\documentclass[12pt]{article}
 \advance\oddsidemargin by -1in
 \advance\evensidemargin
 by -1in
 \marginparsep
 0.4cm
 \advance\topmargin by
 -0.0in
 \makeindex
 \def\Pom{{\bf I\!P}}
 \def\gsim{\mathrel{\rlap{\lower4pt\hbox{\hskip1pt$\sim$}}
 \raise1pt\hbox{$>$}}}

 \newcommand\la{\langle}
 \newcommand\ra{\rangle}
 \newcommand\beq{\begin{equation}}
 \newcommand\noi{\noindent}
 \newcommand\eeq{\end{equation}}
 \newcommand\beqn{\begin{eqnarray}}
 \newcommand\eeqn{\end{eqnarray}}

\begin{document}

\title{\bf Coherence Phenomena\\
in Charmonium Production off Nuclei\\
at the Energies of RHIC and LHC}

\maketitle

\begin{center}

 {\large Boris Kopeliovich$^{1,2,3}$,
Alexander Tarasov$^{1,2,3}$ and J\"org H\"ufner $^{1,4}$}
 \\[1cm]
 $^{1}${\sl Max-Planck Institut f\"ur Kernphysik, Postfach 103980, 69029
Heidelberg, Germany}\\[0.2cm]
 $^{2}${\sl Institut f\"ur Theoretische Physik der Universit\"at, 93040
Regensburg, Germany} \\[0.2cm]
 $^{3}${\sl Joint Institute for Nuclear Research, Dubna, 141980 Moscow
Region, Russia}\\[0.2cm]
 $^{4}${\sl Institut f\"ur Theoretische Physik der Universit\"at,
Philosophenweg 19, 69120 Heidelberg}

\end{center}

\vspace{1cm}
 
\begin{abstract}
\noi
In the energy range of RHIC and LHC the mechanisms of nuclear suppression 
of charmonia are expected to be strikingly different from what is known for
the energy of the SPS. One cannot think any more of charmonium produced 
on a bound nucleon which then attenuates as it passes 
through the rest of the nucleus. The coherence length of charmonium production 
substantially exceeds the nuclear radius in the new energy range. 
Therefore the production
amplitudes on different nucleons, rather than the cross sections, 
add up and interfere, i.e. shadowing is at work. So far no theoretical tool 
has been available to calculate nuclear effects for charmonium production
in this energy regime. We develop a light-cone Green function formalism 
which incorporates the effects of the coherence of the 
production amplitudes and of charmonium wave function formation, and is the
central result of this paper. We found a substantial deviation 
from QCD factorization, namely, gluon shadowing is much stronger for charmonium 
production than it is in DIS. We predict for nuclear effects $x_2$ scaling which
is violated at lower energies by initial state energy loss which must be
also included in order to compare with available data. In this paper only the 
indirect $J/\Psi$s originating from decay of $P$-wave charmonia 
are considered. The calculated $x_F$-dependence of $J/\Psi$ nuclear 
suppression is in a good accord with data. We predict a dramatic variation of
nuclear suppression with $x_F$ in pA and a peculiar peak at $x_F=0$ in
AA collisions at RHIC. 

\end{abstract}


\newpage
 
\vspace*{1cm}

\section{Introduction}

 Charmonium production off nuclei has drawn much attention during the last two
decades since the NA3 experiment at CERN \cite{na3} has found a steep increase
of nuclear suppression with rising Feynman $x_F$. This effect has been
confirmed later in the same energy range \cite{katsanevas}, 
and at higher energy recently by the most precise
experiment E866 at Fermilab \cite{e866}. No unambiguous explanation for these 
observations has been provided yet.
With the advent of RHIC new data are expected soon in the unexplored energy
range. Lacking a satisfactory understanding of nuclear effects for
charmonium production in proton-nucleus collisions it is very difficult to
provide a convincing interpretation of data from heavy ion collisions experiments
\cite{na38,na50} which are aimed to detect the creation of a quark-gluon
plasma using charmonium as a sensitive probe. Many of existing analyses rely
on an oversimplified dynamics of charmonium production which fails to explain
even data for $pA$ collisions, in particular the observed $x_F$ dependence of
$J/\Psi$ suppression. Moreover, sometimes even predictions for 
RHIC employ those simple models.
It is the purpose of present paper to demonstrate that the dynamics of charmonium 
suppression strikingly changes between the SPS and RHIC energies.
We perform full QCD calculations of nuclear effects within framework of 
the light-cone Green function approach aiming to explain observed nuclear effects 
without adjusting any parameters, and to provide realistic predictions for RHIC.

To avoid a confusion, we should make it clear that we will skip 
discussion of any mechanisms
of charmonium suppression caused by the interaction with the produced comoving
matter, although it should be an important effect in central heavy ion
collisions. Instead, we consider suppression which originates from
the production process and propagation of the $\bar cc$ pair through the nucleus. 
It serves as a baseline for search for new physics in heavy ion collisions.

 The present paper is focused on coherence phenomena which are still a rather
small correction for charmonium production at the SPS, 
but whose onset is already observed at
Fermilab  and which are expected to become a dominant effect at the
energies of the RHIC and LHC. One realizes the importance of the coherence effects
treating charmonium production in an intuitive way as a hard 
$\bar cc$ fluctuation which lose coherence with the projectile
ensemble of partons via interaction with the target, and is thus liberated.
In spite of the hardness of the fluctuation, its lifetime in the 
target rest frame increases with energy and eventually exceeds the
nucleus size. Apparently, in this case the $\bar cc$ pair is freed 
via interaction with the whole nucleus, rather than with an individual
bound nucleon as it happens at low energies. Correspondingly, nuclear effects
become stronger at high energies since the fluctuation propagates through the
whole nucleus, and different nucleons compete with each other in freeing
the $\bar cc$. In terms of the conventional
Glauber approach it leads to  shadowing. In terms of the parton model it is analogous
to shadowing of $c$-quarks in the nuclear structure function. 
It turns out (see Sect.~\ref{gluons}) that the 
fluctuations containing gluons in addition to the $\bar cc$ pair are subject
to especially strong shadowing. Since at high energies the weight of such  
fluctuations rises, as well the the fluctuation lifetime, it
becomes the main source of nuclear suppression of open and hidden charm
at high energies, in particular at RHIC. In terms of the parton model
shadowing for such fluctuations containing gluons correspond to gluon shadowing.

The parton model interpretation of charmonium production contains no explicit
coherence effects, but they are hidden in the gluon
distribution function of the nucleus which is supposed to be subject
to QCD factorization. There are, however, a few pitfalls on this way.
First of all, factorization is exact only in the limit of a very hard scale.
That means that one should neglect the effects of the order of the inverse 
$c$-quark mass, in particular the transverse $\bar cc$ separation 
$\la r_T^2\ra \sim 1/m_c^2$. However, shadowing and absorption of $\bar cc$
fluctuations is a source of a strong suppression which is nearly factor of $0.5$ for
heavy nuclei (see Fig.~\ref{e-dep}). QCD factorization misses this effect.
Second of all, according to factorization gluon 
shadowing is supposed to be universal, i.e. one
can borrow it from another process (although we still have no experimental 
information about gluons shadowing, it only can be calculated) and
use to predict nuclear suppression of open or hidden charm. Again, 
factorization turns out to be dramatically 
violated at the scale of charm and gluon
shadowing for charmonium production is much stronger than
it is for open charm or deep-inelastic scattering (DIS) (compare gluon
shadowing exposed in Fig.~\ref{glue-shad} with one calculated in 
\cite{kst2} for DIS). All these important, sometimes dominant effects are missed 
by QCD factorization. This fact once again emphasizes the advantage 
of the light-cone dipole approach which does reproduce QCD factorization
in situations where that is expected to be at work, and it is also able to calculate
the deviations from factorization in a parameter free way.

Unfortunately, none of the existing models for $J/\Psi$ or $\Psi^\prime$
production in $NN$ collisions is fully successful in describing all
the features observed experimentally.
In particular, the $J/\Psi$, $\Psi^\prime$ and $\chi_1$ production cross sections 
in $NN$ collisions come out 
too small by at least an order of magnitude \cite{bhtv}. Only data for production
of $\chi_2$ whose mechanism is rather simple seems to be in good accord with
the theoretical expectation based on the   
color singlet mechanism (CSM) \cite{csm,tv} treating
$\chi_2$ production via glue-glue fusion. The contribution of the color-octet
mechanism is an order of magnitude less that of CSM \cite{tv}, and is even more
suppressed according to \cite{schaefer}. The simplicity of the production
mechanism of $\chi_2$ suggests to use this process as a basis  for the 
study of nuclear effects. Besides, about $40\%$ of the $J/\Psi$s have 
their origin in $\chi$ decays. We drop the subscript of $\chi_2$ in what follows
unless it is important.

\subsection{What has been understood already}

A lot of work has been done already 
and considerable progress has been achieved
in the understanding of many phenomena related to the dynamics of the charmonium 
production and nuclear suppression.
We would like to start with reviewing some of these phenomena which are
employed in present paper.

$\bullet$ 
Relative nuclear suppression of $J/\Psi$ and 
$\Psi^\prime$ has attracted much attention. 
The $\Psi^\prime$ has twice as large radius as the 
$J/\Psi$, therefore should attenuate in nuclear matter much stronger. However,
formation of the wave function
of the charmonia takes time, one cannot instantaneously 
distinguish between these two levels. 
This time interval or so called formation time (length) is 
enlarged at high energy $E_{\Psi}$ by Lorentz time dilation,
 \beq
t_f=\frac{2\,E_{\Psi}}{M_{\Psi^\prime}^2-M_{J/\Psi}^2}\ ,
\label{1}
 \eeq
 and may become comparable to or even longer than the nuclear radius. 
In this case neither $J/\Psi$, nor $\Psi^\prime$ propagates through
the nuclear medium, but a pre-formed $\bar cc$ wave packet \cite{bm}. 
Intuitively, one might even expect an universal nuclear suppression,
indeed supported by data \cite{e772,na38,e866}. However, 
a deeper insight shows that such a point of view
is oversimplified, namely, the mean transverse size of the $\bar cc$ 
wave packet propagating through the nucleus varies depending on the wave 
function of the final meson which the $\bar cc$ is projected to.
In particular, the nodal structure of the $2S$ state substantially enhances
 the yield of $\Psi^\prime$ \cite{kz91,bkmnz} (see in \cite{kh-prl,kh-z} a
complementary interpretation in the hadronic basis). 
 
$\bullet$
Next phenomenon is related to the so called coherence time.
Production of a heavy $\bar cc$ is associated with a 
longitudinal momentum transfer $q_c$ which decreases with energy.
Therefore the production amplitudes on different nucleons add up coherently
and interfere if the production points are within the interval
$l_c=1/q_c$ called coherence length or time,
 \beq
t_c=\frac{2\,E_{\Psi}}{M_{J/\Psi}^2}\ .
\label{2}
 \eeq
This time interval is much shorter than the formation time Eq,~(\ref{1}).
One can also interpret it in terms of the uncertainty principle as the mean
lifetime of a $\bar cc$ fluctuation.
If the coherence time is long compared to the nuclear radius,  
$t_c\gsim R_A$, different nucleons compete with each
 other in producing the charmonium. Therefore, the amplitudes interfere 
destructively leading to an additional suppression called shadowing.
Predicted in \cite{kz91} this effect was confirmed by the NMC
measurements of exclusive $J/\Psi$ photoproduction off nuclei \cite{nmc} 
(see also \cite{bkmnz}). The recent precise data from the HERMES experiment 
\cite{hermes} for electroproduction of
$\rho$ mesons also confirms the strong effect of coherence time \cite{hkn}.

Note that the coherence time Eq.~(\ref{2}) is relevant only for the 
lightest fluctuations $|\bar cc\ra$. Heavier ones which contain additional gluons 
have shorter lifetime. However, at high energies they are also at work 
and become an important source of an extra suppression (see in \cite{kst2} and 
Sect.~\ref{gluons}). They correspond to shadowing of gluons in terms of
parton model. 
In terms of the dual parton model the higher Fock states contain additional 
$\bar qq$ pairs instead of gluons. Their contribution enhances on a nuclear target 
lead to softening of the $x_F$ distribution of the produced charmonium. This mechanism 
has been used in \cite{capella} to explain $x_F$ dependence of charmonium
suppression. However the approach was phenomenological and data were fitted.

The first attempt to implement the coherence time effects into the
dynamics of charmonium production off nuclei has been done in \cite{ya}.
However, the approach still was phenomenological and data also were fitted.
Besides, gluon shadowing (see Sect.~\ref{gluons}) has been missed.

$\bullet$
The total $J/\Psi$-nucleon cross section steeply rises with energy, approximately
as $s^{0.2}$. This behavior is suggested by the observation of a steep 
energy dependence of the cross section of $J/\Psi$ photoproduction at HERA. This fact
goes well along with observation of 
the strong correlation between $x_{Bj}$ dependence of the
proton structure function at small $x_{Bj}$ and the photon virtuality $Q^2$: 
the larger $Q^2$ is (the smaller is its $\bar qq$ fluctuation), 
the steeper the $F_2(x_{Bj},Q^2)$ rises with $1/x_{Bj}$.
Apparently, the cross section of a small size charmonium must rise with energy 
faster than what is known for light hadrons.
The $J/\Psi$-nucleon cross section has been calculated recently in \cite{hikt} 
employing 
the light-cone dipole phenomenology, realistic charmonium wave functions and
phenomenological dipole cross section fitted to data for $F_2(x,Q^2)$ from HERA. 
The results are in a good accord with data for the  
electroproduction cross sections of $J/\Psi$ and $\Psi^\prime$ and also confirm the
steep energy dependence of the charmonium-nucleon cross sections.
Knowledge of these cross sections is very important for understanding 
of nuclear effects in the production of charmonia.
A new important observation made in \cite{hikt} is a strong effect of spin rotation 
associated with boosting the $\bar cc$ system from its rest frame to 
the light cone.  It substantially increases the $J/\Psi$ and especially $\Psi^\prime$
photoproduction cross sections. The effect of spin rotation is also implemented
in our calculations below and it is crucial for restoration of the Landau-Yang 
theorem (see Appendix~C).

$\bullet$
 Initial state energy loss by partons traveling through the nucleus affects
the $x_F$ distribution of produced charmonia \cite{kn1}
especially at medium high energies. A shift in the effective value
of $x_1$, which is the fraction of the incident momentum carried by the produced
charmonium, and the steep $x_1$-dependence 
of the cross section of charmonium production
off a nucleon lead to a dramatic nuclear 
suppression at large $x_1$ (or $x_F$) in a good
agreement with data \cite{na3,katsanevas}. The recent analyses \cite{eloss,eloss1}
of data from the E772 experiment for Drell-Yan process on nuclei 
reveals for the first time a nonzero and rather large energy loss.

\subsection{Outline of the paper}

The paper is organized as follows.
The light-cone (LC) formalism for 
gluon-nucleon collision formulated in the rest frame of the target nucleon
is introduced in Sect.~\ref{LC}. As usual, the amplitude is represented as a
convolution in the impact parameter space of the LC wave functions
of the incident gluon and the final charmonium, and the dipole transition
cross section. The latter corresponds to the interaction with the target
which causes a transition between $|\bar cc\ra_G$ and $|\bar cc\ra_\chi$, 
which are the
charm-anticharm pairs with quantum numbers of a gluon and a $\chi$
respectively. The transition amplitude $\Sigma^{tr}$ can be expressed in
terms of the ordinary flavor independent dipole cross section $\sigma_{\bar
qq}$ of interaction of a colorless $\bar qq$ dipole with a nucleon which is
rather well known from phenomenology.
 
 While the perturbative gluon wave function, an analog to the photon one,
is known, the LC wave function of a
charmonium needs to be constructed. Even if the  wave function in the
rest frame of the charmonium is known, it is not a straightforward procedure
to boost it to the infinite momentum frame. We apply the widely accepted
prescription for the Lorentz boost, which is rather successful in describing
data for exclusive photoproduction of charmonia \cite{hikt}. It is presented
in Appendix~A for the case of the $\chi$ wave function. One may not expect
important relativistic effects for such a heavy system as a
charmonium. Indeed, it is demonstrated in Appendix~B that the distribution
over fraction of the $\chi$ momentum carried by the quarks peaks at $1/2$
with a tiny width. However, the Melosh spin rotation effect is still very
important. It is demonstrated in \cite{hikt} how much it affects the
photoproduction cross section of $J/\Psi$ and especially $\Psi^\prime$. In
the case of $\chi$ gluoproduction inclusion of the spin rotation restores
the Landau-Yang theorem which forbids production of $|\chi_1\ra$ by an 
on-mass-shell gluon. Otherwise, it would be badly violated even in the
nonrelativistic limit where one might expect the spin rotation effects to
vanish. What is rather obvious in the parton model, still 
needs special efforts to be
proven within the LC approach. The fine tuning leading to cancelation of
different parts of the $\chi_1$ production amplitude is demonstrated in
Appendix~C. This can be considered as another strong 
support for the Lorentz boost procedure we are using.
 
 The production of charmonia off nuclei is controlled by few length scales
as is discussed above and in Sect.~\ref{quarks}. 
Once a colorless $\bar cc$ wave packet
is produced, it is evolving during propagation through the nucleus
due to absorption and transverse motion of the quarks. The result should
be projected to the charmonium wave
function. The length scale of the evolution is controlled by the formation
time Eq.~(\ref{1}).
 
 Another much shorter length scale is controlled 
by the coherence time Eq.~(\ref{2}). At high energy it becomes long 
leading to shadowing which enhances the suppression of charmonia.
This is the new phenomenon which is not yet at work at the SPS, 
but is expected to nearly saturate at RHIC.
 
 We start with the
effects of quantum coherence for simple limiting situations. In
Sect.~\ref{long-lc} we study the case of a very long $l_f\gg l_c\gg R_A$ when
the transverse quark separation is frozen by the Lorentz time dilation for
the time of propagation through the nucleus. While the final state absorption
occurs with the conventional dipole cross section $\sigma_{\bar qq}(r_T)$, it
is not obvious which cross section controls shadowing. It is the key
observation of Sect.~\ref{long-lc} that this is the dipole cross section
$\sigma_3$ of interaction of a colorless system consisting of three partons,
$|\bar qqG\ra$. The combined effect of shadowing and absorption is given by
Eq.~(\ref{1.29}) - (\ref{1.30}).
 
 The case of a not too long coherence time
$l_c\sim R_A$ is considered in
 Sect.~\ref{short-lc}. Since $l_f\gg l_c$, the
time taken by the produced $\bar cc$ for its further development can be
rather long, $l_f\gg R_A$ and the transverse size of the $\bar cc$ pair can
be treated as frozen.
 Eq.~(\ref{2.1}) interpolating between the regimes of
very short and long coherence lengths is governed by the longitudinal
momentum transfer $q_c=1/l_c$.
 
 In the general case considered in
Sect.~\ref{generalcase} any of $l_c$ and $l_f$ can be either short or long and
one must take care of the transverse size fluctuations of a color
dipole propagating through the nucleus. The appropriate approach is the path
integral technique summing up all possible paths of the partons
\cite{kz91}. Evolution of a $\bar cc$ wave packet in nuclear medium is
described by the LC Green functions satisfying the two dimensional 
Schr\"odinger equations
(\ref{3.5}), which are different for colorless and colored dipoles. The
central result of this paper, the amplitude of the process $G\,A \to \chi\,X$
is described by Eq.~(\ref{3.2}) which is illustrated pictorially in
Fig.~\ref{general-fig}.
 
 Evaluation of the effects of coherence and formation
is performed in Sect.~\ref{numerics} and the results are demonstrated in
Fig.~\ref{e-dep}. The nuclear transparency first rises with energy due to the
formation length effects (color transparency), but then, falls down at
higher energies due to growing coherence length.

In Sect.~\ref{gluons} a 
third scale governing the nuclear effects is introduced. It is related to
the higher Fock components in the projectile gluon, $|\bar cc\,G\ra$,
$|\bar cc\,2G\ra$ etc. Since the gluons usually carry a small fraction of
the total momentum, these fluctuations are rather heavy compared to
$|\bar cc\ra$, hence they have a coherence time shorter than in Eq.~(\ref{2}).
Therefore, the contribution of these fluctuation to nuclear shadowing
which must be associated with gluon shadowing, is delayed down 
to smaller values of $x_2$. 

First of all one must develop an impact parameter approach for fluctuations 
containing gluons. It is quite a difficult task, but it pays off when one needs
to calculate shadowing which is given by a simple eikonalization,
or by employing the Green function technique if $x_2$ is not small enough.
This is done in Appendix~D and the results are summarized in Sect.~\ref{lc-wf}.
It turns out that the twelve Feynman graphs for the LC wave function of
the $|\bar ccG\ra$ depicted in Fig.~\ref{graphs}
and calculated in Appendix~D are reduced to 
the main contribution which corresponds to the production of $\chi$ directly
from the projectile gluon prior or after the interaction with the target,
as is illustrated in Fig.~\ref{dy}. Therefore the dipole formalism appropriate for
gluon shadowing is pretty similar (up to color factors) to that for Drell-Yan 
reaction \cite{hir,bhq,kst1,krt3}. 
The LC wave function of the $|\bar ccG\ra$
state where the $\bar cc$ pair is colorless, turns out to be very different from
that for the $|\bar qqG\ra$ fluctuations in DIS where the strong nonperturbative 
interaction between the color-octet $\bar qq$ and the gluon substantially 
diminishes the $\bar qq$ -- glue separation and reduces shadowing of gluons.
Indeed, gluon shadowing for $\chi$ production depicted in 
Fig.~\ref{gluon-shadowing} is much stronger than calculated for DIS in
\cite{kst2}.

There are some indications that gluons in nuclei may be enhanced at 
at $x_2\sim 0.1$. This effect increasing the production rate of charmonia 
at small$x_F$ is discussed and estimated in Sect.~\ref{antishadowing}

More effects are to be
included in order to compare with data. In Sect.~\ref{eloss} the corrections
for energy loss by the projectile partons which are due to initial state
interactions are evaluated according to the prescription of \cite{kn1,eloss,eloss1}.
This effect dramatically violates $x_2$ scaling. While it is the main mechanism
shaping the $x_F$ dependence of charmonium suppression at medium high energies
(the SPS energy range), its influence is essentially reduced at Fermilab, 
and no visible energy loss corrections are expected at RHIC.

Eventually, in order to compare with available data for $J/\Psi$ production
off nuclei one has to make corrections for decay $\chi \to J/\Psi\,\gamma$. 
This is done in Sect.~\ref{decay} and the corrections are found to be rather small. 
Indeed, they are basically the same in pN and pA collisions and essentially 
cancel out in the ratio.

 Incorporating all these effects in our 
calculations in Sect.~\ref{results} we arrive at quite a good agreement
with available data. While energy loss is the dominant effect at $200\,GeV$
(Fig.~\ref{na3}), the
steep $x_F$ dependence of nuclear suppression at $800\,GeV$ (Fig.~\ref{xf-dep})
is a combined effect of quark and gluon shadowing and energy loss.
This is the first manifestation of the onset of gluon shadowing,
but it is expected to to become the main source of charmonium 
suppression at RHIC. Our predictions depicted in Fig.~\ref{rhic} demonstrate
a dramatic variation of nuclear suppression versus small $x_F$ for $pA$ collisions.
Translated into suppression in $AA$ collisions the same effects form a peculiar narrow 
peak at $x_F=0$.

Our conclusions and outlook are presented in Sect.~\ref{conclusions}.
This is the first full QCD calculation of coherence effects in charmonium 
production off nuclei demonstrating that coherence is the dominant phenomenon 
which governs nuclear suppression of charmonia at the energies of RHIC and LHC.

\section{The light-cone dipole formalism for charmonium production off
a nucleon}\label{LC}
 
 The important advantage of the light-cone (LC) dipole approach is
its simplicity in the calculations of nuclear effects. It has been suggested two
decades ago \cite{zkl} that quark configurations (dipoles) with fixed
transverse separations are the eigenstates of interaction in QCD. Therefore
the amplitude of interaction with a nucleon is subject to eikonalization in
the case of a nuclear target. In this way one effectively sums the Gribov's
inelastic corrections in all orders.
 
\begin{figure}[tbh]
\includegraphics{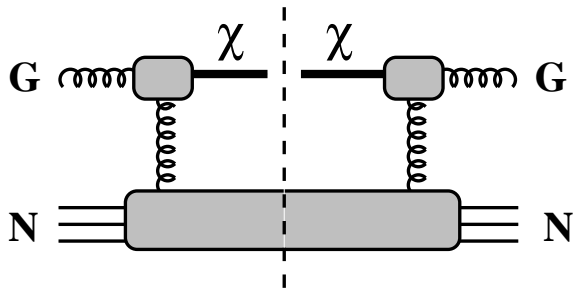}
\begin{center}
\vspace{4cm}
\parbox{13cm}
 {\caption[Delta]
 {\sl Perturbative QCD mechanism of
production of the $\chi$ states
 in a gluon-nucleon collision.}
\label{fig1}}
\end{center}
\end{figure}

 Assuming that the produced $\bar cc$ pair is sufficiently small so that
multigluon vertices can be neglected we can write the cross section for
$G\,N\to \chi\,X)$ as (see Fig.~\ref{fig1}),
 \beq
 \sigma(GN\to\chi X)=\frac{\pi}{2(N_c^2-1)}\,
 \sum\limits_{a,b}
\int \frac{d^2k_T}{k_T^4}\,
 \alpha_s(k_T^2)\,{\cal F}(x,k_T^2)\,
\Bigl|M_{ab}(\vec k_T)\Bigr|^2\ ,
\label{1.0}
 \eeq
 where ${\cal F}(x,k_T^2)=\partial G(x,k_T^2)/\partial (\ln k_T^2)$
 is
the unintegrated gluon density, $G(x,k_T^2)=x\,g(x,k_T^2)$
($x=M_{\chi}^2/\hat s$); $M_{ab}(\vec k_T)$ is the fusion amplitude
$G\,G\to\chi$ with $a,\ b$ being the gluonic indexes.

 In the rest frame of the nucleon the amplitude can be represented in
terms of the $\bar cc$ LC wave functions of the projectile gluon and
 ejectile
charmonium,
 \beq
 M_{ab}(\vec k_T) = \frac{\delta_{ab}}{\sqrt{6}}\,
\int\limits_0^1 d\alpha\,\int d^2 r_T\,
 \sum\limits_{\bar\mu\mu}\,
\Bigl(\Phi_{\chi}^{\bar\mu\mu}(\vec r_T,\alpha)\Bigr)^*\,
 \left[e^{i\vec
k_T\cdot\vec r_1} -
 e^{i\vec k_T\cdot\vec r_2}\right]\,
\Phi_{G}^{\bar\mu\mu}(\vec r_T,\alpha)\ ,
\label{1.00}
\eeq
 where
 \beq
 \vec r_1=(1-\alpha)\,\vec r_T,\ \ \ \ \ \
 \vec r_2=
 - \vec\alpha\,\vec r_T\ .
\label{1.00a}
\eeq
 
 For the sake of simplicity we separate the normalization color
factors
 $\Bigl\la\bar cc,{\{8\}_a}\Bigr|$ and $\la\bar cc,\{1\}|$ from the LC
wave
 function of the gluon and charmonium respectively, and calculate the
matrix element,
 \beq
 \Bigl\la\bar cc,{\{8\}_a}\Bigr|
{1\over2}\,\lambda_b\Bigl|\bar cc,\{1\}\Bigr\ra =
\frac{\delta_{ab}}{\sqrt{6}}\ ,
\label{1.01}
\eeq
 which is shown explicitly in (\ref{1.00}). Thus, the functions
 $\Phi_{G(\chi)}^{\bar\mu\mu}(\vec r_T,\alpha)$ in (\ref{1.00}) represent
 only the spin- and coordinate dependent parts of the corresponding full
 wave
 functions.
 
 The gluon wave function differs only by a factor from the
 photon one,
 \beq
 \Phi_{G}^{\bar\mu\mu}(\vec r_T,\alpha) =
 \frac{\sqrt{2\,\alpha_s}}{4\pi}\,
 \Bigl(\xi^\mu_c\Bigr)^\dagger\,\hat O\,
 \tilde\xi^{\bar\mu}_{\bar c}\,
 K_0(\epsilon\,r_T)\ ,
\label{1.02}
\eeq
 where $\xi^{\mu}_{c}$ is the $c$-quark spinor, and
 \beqn
 \tilde\xi^{\bar\mu}_{\bar c} &=&
 i\,\sigma_y\,\xi^{\bar\mu}_{\bar
 c}\Bigr.^*\ ,
 \label{1.02a}\\
 \hat O &=& m_c\,\vec\sigma\cdot\vec e +
 i(1-2\alpha)\,(\vec\sigma\cdot\vec n)\,
 (\vec e\cdot\vec\nabla) +
 (\vec
 e\times\vec n)\vec\nabla\ ,
\label{1.02b}\\
 \epsilon^2 &=& Q^2\alpha(1-\alpha)+m_c^2\ ,
 \label{1.02c}\\
 \vec\nabla &=& \frac{d}{d\,\vec r_T}\ .\nonumber
\label{1.03}
 \eeqn
 The gluon has virtuality $Q^2$ and polarization vector $\vec e$
 and is moving along the unit vector $\vec n$ (in what follows we consider
 only transversely polarized gluons, $\vec e\cdot\vec n=0$).
 
 The expression for
 the LC wave function of a charmonium related by the
 Lorentz boost from the
 charmonium rest frame is rather complicated and is
 moved to Appendix A. This
 complexity is a consequence of the nonlocal
 relation between the LC
 variables ($\vec r_T$, $\alpha$) and the
 components of the 3-dimensional
 relative $\bar cc$ radius-vector $\vec r$
 in the rest frame of the
 charmonium. Also the Melosh spin rotation leads
 to a nontrivial relations
 between the two wave functions (see {\it e.g.}
 in \cite{hikt}). This is a
 relativistic effect, it vanishes in the limit
 of small velocity $v\to 0$ of
 the quarks in the charmonium rest
 frame. 

A word of caution is in
 order. In some cases the Melosh
 spin rotation is important even in the limit
 of vanishing quark velocity
 $v\to 0$. An example is the Landau-Yang theorem
 \cite{landau} which
 forbids production of the $\chi_1$ state by two massless
 gluons. However,
 the LC approach leads to creation of the $\chi_1$ even in
 the limit $v\to
 0$ if the effect of spin rotation is neglected. It is
 demonstrated in
 Appendix B that the Landau-Yang theorem is restored only if
 the Melosh
 spin rotation is included. Such a cancelation of large values is
 a kind
 of fine tuning and is a good support for the procedure of Lorentz
 boosting which we apply to the charmonium wave functions.
 
 Since the
 gluon LC wave function smoothly depends on $\alpha$
 while the charmonium
 wave function peaks at $\alpha=1/2$ with a tiny width
 estimated in Appendix
 B, $\la(\alpha-1/2)^2\ra=0.01$,
 we can replace the charmonium wave function
 in the matrix element in
 (\ref{1.00}) with
 \beq
 \Phi^{\bar\mu\mu}_{\chi}(r_T,\alpha) \approx
 \delta\left(\alpha-{1\over2}\right)\,
 \int
 d\alpha\,\Phi^{\bar\mu\mu}_{\chi}(r_T,\alpha)\ .
\label{1.06}
\eeq

 It is convenient to expand the LC charmonium wave function in
 powers
 of $v$. The result depends on the total momentum $J$ and its
 projection $J_z$ on the direction $\vec n$.
 The charmonium wave function
 integrated over $\alpha$ has
 the form,
 \beq
 \int d\alpha\,
 \Phi^{\bar\mu\mu}_{\chi}(r_T,\alpha) =
 \Bigl(\xi^\mu\Bigr)^{\dagger}\,
 \left[\vec\sigma\cdot\vec e_{\pm} +
 {1\over m}\,(\vec e_{\pm}\times\vec
 n)\cdot\vec\nabla -
 \frac{1}{2\,m_c^2}\,(\vec e_{\pm}\cdot\vec\nabla)\,
 (\vec\sigma\cdot\vec\nabla)\right]\,
 \tilde\xi^{\bar\mu}\,W +
 O(v^4)\ ,
\label{1.07}
\eeq
 where
 \beq
 W=\frac{\vec e_{\pm}\cdot\vec r_T}{r_T}\,
 \left[R(r_T) + \frac{3}{4\,m_c^2}\,R^{\prime\prime}(r_T) +
 O(v^4)\right]\
 ,
\label{1.08}
\eeq
 and $R(r)$ is the radial part of the P-wave charmonium
 in its rest
frame (see derivation in Appendix A).
 The new notations for the polarization
vectors are,
 \beqn
 \vec e_+ &=& - \frac{\vec e_x+i\vec e_y}
{\sqrt{2}}\ ,\nonumber\\
 \vec e_- &=& \frac{\vec e_x-i\vec e_y}
 {\sqrt{2}}\
.
\label{1.09}
\eeqn
 
 In what follows we use the LC wave functions of gluons and
charmonium in order to calculate matrix elements of operators which
 depend
only on the LC variables $r_T$ and $\alpha$. Therefore,
 for the sake of
simplicity we can drop off the indexes
 $\mu,\bar\mu$ and summation over them,
{\it i.e.} replace
 \beq
\sum\limits_{\mu\bar\mu}\Bigl(\Phi^{\mu\bar\mu}_{\chi}
 (\vec
r_T,\alpha)\Bigr)^*\,
 \Phi^{\mu\bar\mu}_{G}(\vec r_T,\alpha) \Rightarrow
\Phi^*_{\chi}(\vec r_T,\alpha)\,
 \Phi_{G}(\vec r_T,\alpha)
\label{1.10}
\eeq
 With this convention we can rewrite the cross section
 Eq.~(\ref{1.00}) as,
 \beqn
 && \sigma(GN\to\chi X) =
 \int\limits_0^1
 d\alpha
 \int\limits_0^1 d\alpha'
 \int d^2r_T\,d^2r_T^{\prime}\nonumber\\
 &\times& \Bigl\{\Phi_{\chi}^*(\vec r_T,\alpha)\,
 \Phi_{\chi}(\vec
 r_T^{\,\prime},\alpha^{\prime})\,
 \Sigma^{tr}(\vec r_T,\vec
 r_T^{\prime},\alpha,\alpha^{\prime})\,
 \Phi_{G}(\vec r_T,\alpha)\,
 \Phi_{G}^*(\vec r_T^{\,\prime},\alpha^{\prime})\Bigr\}\ ,
\label{1.1}
\eeqn
 where the transition cross section $\Sigma^{tr}$ is a
 combination of
 dipole cross sections,
 \beq
 \Sigma^{tr}(\vec r_T,\vec
 r_T^{\,\prime},\alpha,\alpha^{\prime})
 = {1\over16}\,\Bigl[\sigma_{\bar
 qq}(\vec r_1-\vec
 r_2^{\,\prime})+
 \sigma_{\bar qq}(\vec r_2-\vec
 r_1^{\,\prime}) -
 \sigma_{\bar qq}(\vec r_1-\vec
 r_1^{\,\prime}) -
 \sigma_{\bar qq}(\vec r_2-\vec r_2^{\,
 \prime})\Bigr]\ ,
\label{1.2}
 \eeq
 and $\vec r_1,\ \vec r_2'\ \vec r_1^\prime$ and $\vec r_2^\prime$ are
 defined like in Eq.~(\ref{1.00a}). The dipole cross section \cite{zkl},
 \beq
 \sigma_{\bar qq}(r_T,s) =
 \frac{4\pi}{3}\int
 \frac{d^2
 k_T}{k_T^4}\,
 \alpha_s(k_T^2)\,{\cal F}(x,k_T^2)\,
 \left(1-e^{i\vec
 k_T\cdot\vec r_T}\right)\ ,
\label{1.3a}
  \eeq
 corresponds to interaction of a colorless $\bar qq$ pair of
 transverse
 separation $r_T$ with a nucleon at the squared c.m. energy $s$
 and $x=4k_T^2/s$. The explicit $s$-dependence of the cross sections is
 dropped in Eqs.~(\ref{1.1}), (\ref{1.2}).
 
 Since small distances $r_T\sim
 r_T^{\,\prime}\sim 1/m_c$ dominate in the
 integral in Eq.~(\ref{1.1}) one
 can make use of the approximation
 \beq
 \sigma_{\bar qq}(r_T)\Bigr|_{r_T\to
 0} = C(s)\,r_T^2\ ,
\label{dipole}
\eeq
 then $\Sigma^{tr}$ reduces to the very simple form,
 \beq
 \Sigma^{tr}(\vec r_T,\vec
 r_T^{\,\prime},\alpha,\alpha^{\prime})=
{C(s)\over8}\,(\vec r_T\cdot\vec r_T^{\,\prime}) =
\sum\limits_{\lambda=1,2}
 (\vec e^{\,\lambda}_t\cdot\vec d)\,
 (\vec
e^{\,\lambda}_t\cdot\vec d^{\,\prime})^*\ ,
\label{1.4}
\eeq
 where
 \beq
 \vec d=\sqrt{C(s)\over8}\,\vec r_T\ ,\ \ \ \
 \vec
 d^{\,\prime}=\sqrt{C(s)\over8}\,\vec r_T^{\,\prime}\ .
\label{1.5}
 \eeq
The vectors $\vec e^{\,\lambda}_t$ can be interpreted as a
 polarization vector
 of the Weizs\"acker-Williams gluon of the
 target.
 
 Within this
 approximation the cross section (\ref{1.1}) can
 be represented in the
 form,
 \beq
 \sigma(GN\to \chi X) = \sum\limits_{\lambda}\,
 \Bigl|A^{(\lambda)}\Bigr|^2\ ,
\label{1.6}
\eeq
 where
 \beq
 A^{(\lambda)}=
 \int\limits_0^1 d\alpha
 \int
 d^2r_T\,\Phi^*_{\chi}(\vec r_T,\alpha)\,
 (\vec e^{\lambda}_T\cdot \vec
 d)\,
 \Phi_G(\vec r_T,\alpha)\ .
\label{1.7}
\eeq
 
\section{Production of charmonia off nuclei: shadowing of 
\boldmath$c$-quarks}\label{quarks}

Nuclear effects in the production of a $\chi$ are controlled by the
coherence and formation lengths which are defined in (\ref{1}), (\ref{2}).
One can identify two limiting cases. The first one corresponds to the
situation where both $l_c$ and $l_f$ are shorter that the mean spacing 
between
bound nucleons. In this case one can treat the process classically, the
charmonium is produced on one nucleon inside the nucleus and attenuates
exponentially with an absorptive cross section which is the inelastic
$\chi-N$ one. This simplest case is described in \cite{kn1,gh}.
 
In the
limit of a very long coherence length $l_c\gg R_A$ one can think about a 
$\bar cc$
fluctuation which emerges inside the incident hadron long before the 
interaction with the nucleus. Different bound nucleons compete and shadow
each other in the process of liberation of this fluctuation. This causes an
additional attenuation in addition to inelastic collisions of the
produced
color-singlet $\bar cc$ pair on its way out of the nucleus. Since
$l_c \ll
l_f$ an intermediate case is also possible where $l_c$ is shorter
than the
mean internucleon separation, while $l_f$ is of the order or
longer than the nuclear radius.
 
\subsection{The high energy limit, \boldmath$l_c \gg
R_A$}\label{long-lc}
 
\begin{figure}[tbh]
 \includegraphics{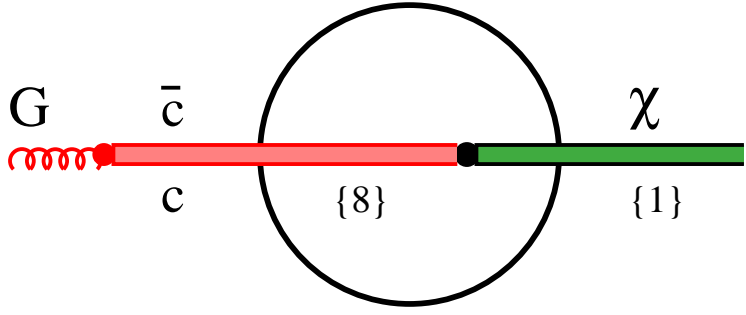}
\begin{center}
\vspace{4.5cm}
\parbox{13cm}
 {\caption[Delta]
 {\sl The incident gluon converts onto a
$\bar cc$ pair long in advance of
 the nucleus. The pair propagates and
attenuates with the absorption cross
 section $\sigma_3(r_T)$ (see the text)
up to the point where it converts
 into a colorless $\bar cc$ pair with
quantum numbers of $\chi$. Then it
continues propagating through the nucleus, and is
attenuated with the cross section
 $\sigma_{\bar qq}(r_T)$.}
\label{highenergy}}
\end{center}
 \end{figure}

 We start with the case of very long 
 coherence and formation lengths, $l_c,l_f\gg R_A$. In this
case the
 incident gluon converts into a color-octet $\bar cc$ fluctuation
with a
 lifetime much longer than the nuclear radius. Therefore one can
consider
 this fluctuation as produced long in advance and propagating through
the
 whole nucleus as is illustrated in Fig.~\ref{highenergy}.
 Then we can employ the results of Ref.~\cite{zk} for the
evolution of the density matrix a $\bar qq$ ($\bar cc$ in
 our case) wave
packet propagating through the nuclear medium,
 \beq
 R(\vec r_T,\vec
r_T^{\,\prime},\alpha,\alpha^\prime|z) =
 R^{(1)}(\vec r_T,\vec
r_T^{\,\prime},\alpha,\alpha^\prime|z)\,\hat P_1+
 {1\over8}\,R^{(8)}(\vec
r_T,\vec
 r_T^{\,\prime},\alpha,\alpha^\prime|z)\,\hat P_8\ ,
\label{1.10a}
 \eeq
 where $R(\vec r_T,\vec r_T^{\,\prime},
 \alpha,\alpha^\prime|z)$ is
 the density matrix of the $\bar
 cc$ propagating through nuclear medium which
 depends on
 the transverse separation $\vec r_T$, fraction $\alpha$ of the LC
 momentum, and on the longitudinal coordinate $z$; $\hat P_1$ and
 $\hat P_8$ are
 the projection operators on the singlet and
 octet states of the $\bar qq$,
 respectively, and satisfy the
 relations,
 \beqn
 && \hat P_1+\hat P_8=1\ ;
 \nonumber\\
 &&Tr\,\hat P_1=1\ ;\ \ \ \ Tr\,\hat P_8=8\ .
\label{1.11}
\eeqn
 The matrices $R^{(1)}$ and $R^{(8)}$ are the solutions of
 the evolution 
equations,
 \beqn
 \frac{d\,R^{(1)}(\vec r_T,\vec
r_T^{\,\prime},\alpha,\alpha^\prime|z)}{d\,z}&=&
\rho_A(z)\Bigl[-{1\over2}\,\Sigma^{(1)}\,
 R^{(1)}(\vec r_T,\vec
r_T^{\,\prime},\alpha,\alpha^\prime|z)
 \nonumber\\
 &+& \Sigma^{(tr)}\,
R^{(8)}(\vec r_T,\vec r_T^{\,\prime},\alpha,\alpha^\prime|z)
 \Bigr]\ ,
\label{1.12}
\eeqn
 \beqn
 \frac{d\,R^{(8)}(\vec r_T,\vec
 r_T^{\,\prime},\alpha,\alpha^\prime|z)}{d\,z}&=&
 \rho_A(z)\Bigl[8\,\Sigma^{(tr)}\,
 R^{(1)}(\vec r_T,\vec
 r_T^{\,\prime},\alpha,\alpha^\prime|z)
 \nonumber\\
 &-& \Sigma^{(8)}\,
 R^{(8)}(\vec r_T,\vec r_T^{\,\prime},\alpha,\alpha^\prime|z)
 \Bigr]\ ,
\label{1.13}
\eeqn
 where the nuclear density $\rho_A(z)$ depends on
 the longitudinal
 coordinate $z$ and (implicitly) the impact parameter
 $b$. The transition cross
 section $\Sigma^{tr}$ is defined
 in (\ref{1.2}).
 
 The function
 \beq
 \Sigma^{(1)}\equiv \Sigma^{(1)}(\vec r_T,\vec
 r_T^{\,\prime},\alpha,\alpha^\prime) =
 \sigma_{\bar qq}(\vec r_T) +
 \sigma_{\bar qq}(\vec r_T^{\,\prime})
\label{1.14}
 \eeq
 is the total cross section for the interaction of a 4-quark ensemble,
 two color-singlet $\bar cc$ pairs, with a nucleon.
 Correspondingly,
 \beqn
 \Sigma^{(8)} &\equiv& \Sigma^{(8)}(\vec r_T,\vec
 r_T^{\,\prime},\alpha,\alpha^\prime) =
 {1\over8}\,\Bigl[2\,
 \sigma_{\bar
 qq}(\vec r_1-\vec r_2^{\,\prime}) +
 2\,\sigma_{\bar qq}(\vec r_2-\vec
 r_1^{\,\prime})\nonumber\\
 &+& 7\,\sigma_{\bar qq}(\vec r_1-\vec
 r_1^{\,\prime}) +
 7\,\sigma_{\bar qq}(\vec r_2-\vec r_2^{\,\prime}) -
 \sigma_{\bar qq}(\vec r_1-\vec r_2) -
 \sigma_{\bar qq}(\vec
 r_1^{\,\prime}-\vec
 r_2^{\,\prime})\Bigr]\ ,
\label{1.15}
 \eeqn
 is the total cross section, where a
 4-quark system consisting of two color-octet $\bar cc$
 pairs whose centers
 of gravity coincide, interact with a nucleon.
 
 The initial conditions for a $\bar cc$ pair
 originating from
 the projectile gluon are,
 \beqn
 R^{(1)}(\vec r_T,\vec
 r_T^{\,\prime},\alpha,\alpha^\prime|z)\Bigr|_{z\to-\infty}
 &=& 0\ ,
\label{1.16}\\
R^{(8)}(\vec r_T,\vec
r_T^{\,\prime},\alpha,\alpha^\prime|z)\Bigr|_{z\to-\infty}
 &=& \Phi_G(\vec
r_T,\alpha)\,
 \Phi^*_G(\vec r_T^{\,\prime},\alpha^\prime)\ .
\label{1.17}
 \eeqn
 
 With the evolution equations (\ref{1.12}), (\ref{1.13}) we are in
position
 to extend Eq.~(\ref{1.1}) to the case of a nuclear target,
 \beqn
\sigma(GA\to\chi X) &=& \int d^2b
 \int\limits_0^1 d\alpha
 \int\limits_0^1
d\alpha'
 \int d^2r_T\,d^2r_T^{\prime}\,
 \Phi_{\chi}^*(\vec r_T,\alpha)\,
\Phi_{\chi}(\vec r_T^{\,\prime},\alpha^{\prime})\,
 \nonumber\\ &\times&
R^{(1)}(\vec r_T,\vec
 r_T^{\prime},\alpha,\alpha^{\prime}|z_+)\ ,
\label{1.18}
 \eeqn
 where $z_+\to\infty$, and $R^{(1)}$ implicitly depends on the impact
 parameter $b$. This expression includes all the multiple color exchanges
 between
 the projectile $\bar cc$ pair and bound nucleons in the target
 which
 eventually convert the initial color octet $\bar cc$ into the final
 colorless state we are interested in.
 
 Color exchanges on different
 nucleons add up incoherently, and the cross
 section is rather small due to
 smallness of the mean transverse separation
 for the heavy $\bar cc$
 pair. Therefore, we keep only the lowest (first)
 order in $\Sigma^{tr}$, but
 all higher orders in $\Sigma^{8}$ and
 $\Sigma^{1}$. Then, the cross section
 of the process $G\,A \to \chi\,X$
 reads,
 \beqn
 && \sigma(GA\to\chi X) =
 \int d^2b\,
 \int\limits_0^1 d\alpha
 \int\limits_0^1 d\alpha'
 \int
 d^2r_T\,d^2r_T^{\prime}\nonumber\\
 &\times& \Bigl\{\Phi_{\chi}^*(\vec
 r_T,\alpha)\,
 \Phi_{\chi}(\vec r_T^{\,\prime},\alpha^{\prime})\,
 \widetilde\Sigma(\vec r_T,\vec
 r_T^{\prime},\alpha,\alpha^{\prime})\,
 \Phi_{G}(\vec r_T,\alpha)\,
 \Phi_{G}^*(\vec
 r_T^{\,\prime},\alpha^{\prime})\Bigr\}\ ,
\label{1.19}
 \eeqn
 where
 \beqn
 && \widetilde\Sigma(\vec r_T,\vec
 r_T^{\,\prime},\alpha,\alpha^{\prime}) =
 \frac{2\,\Sigma^{(tr)}(\vec
 r_T,\vec
 r_T^{\,\prime},\alpha,\alpha^{\prime})}{
 \Sigma^{(1)}(\vec
 r_T,\vec
 r_T^{\,\prime},\alpha,\alpha^{\prime}) -
 \Sigma^{(8)}(\vec
 r_T,\vec
 r_T^{\,\prime},\alpha,\alpha^{\prime})}\nonumber\\
 &\times&
 \left\{
 \exp\left[-{1\over2}\,
 \Sigma^{(8)}(\vec r_T,\vec
 r_T^{\,\prime},\alpha,\alpha^{\prime})\,T(b)\right] -
 \exp\left[-{1\over2}\,
 \Sigma^{(1)}(\vec r_T,\vec
 r_T^{\,\prime},\alpha,\alpha^{\prime})\,T(b)\right]\right\}\ .
\label{1.20}
\eeqn
 Here
 \beq
 T(b) = \int\limits_{-\infty}^{\infty} dz\,\rho_A(b,z)\
 ,
\label{1.21}
\eeq
 is the nuclear thickness function.
 
 Eq.~(\ref{1.20}) can be modified to a form which makes its
 physical meaning more transparent (for
 the sake of brevity we drop the
 variables $\vec r_T,\ \vec r_T^{\,\prime},\
 \alpha,\ \alpha^\prime$ in
 functions $\Sigma^{tr},\ \Sigma^{(1)}$ and
 $\Sigma^{(8)}$),
 \beqn
 \widetilde\Sigma(\vec r_T,\vec
 r_T^{\,\prime},\alpha,\alpha^{\prime}) &=&
 \int
 d^2b\int\limits_{-\infty}^{\infty} dz\
 \rho_A(b,z)\,\exp\left[-{1\over2}\,
 \Sigma^{(1)}\,T_+(b,z)\right]\nonumber\\
 &\times& \Sigma^{(tr)}\,
 \exp\left[-{1\over2}\,\Sigma^{(8)}\,T_-(b,z)\right]\ ,
\label{1.22}
\eeqn
 where
 \beqn
 T_+(b,z)=\int\limits_{z}^{\infty} dz^\prime
 \rho_A(b,z^\prime)\nonumber\\
 T_-(b,z)=\int\limits_{-\infty}^{z} dz^\prime
 \rho_A(b,z^\prime)\nonumber\\
\label{1.23}
\eeqn
 
 Furthermore, we introduce the following notations,
 \beq
\sigma_3(r_T,\alpha)=
 \Sigma^{(8)}(\vec r_T,\vec
r_T^{\,\prime},\alpha,\alpha^{\prime})\Bigl|_{r_T^\prime=0}=
{9\over8}\,\Bigl[\sigma_{\bar qq}(\alpha r_T)+
 \sigma_{\bar
qq}[(1-\alpha)r_T]\Bigr]-
 {1\over8}\,\sigma_{\bar qq}(r_T)\ ;
\label{1.24}
\eeq
 \beq
 \Delta\Sigma^{(8)} (\vec r_T,\vec
r_T^{\,\prime},\alpha,\alpha^{\prime}) = \Sigma^{(8)}(\vec
 r_T,\vec
r_T^{\,\prime},\alpha,\alpha^{\prime})- \sigma_3(\vec
 r_T, \alpha) -
\sigma_3(\vec r_T^{\,\prime},\alpha^{\prime})
\label{1.25}
\eeq
 
 Since heavy quarks are expected to be produced with a small
separation, we can employ the approximation
 (\ref{dipole}),
 \beq
\Delta\Sigma^{(8)} (\vec r_T,\vec
 r_T^{\,\prime},\alpha,\alpha^{\prime}) =
- {1\over4}\,\Bigl[7-9(\alpha+\alpha^\prime)
 +
18\,\alpha\,\alpha^\prime\Bigr]\
 \vec r_T\cdot\vec r_T^{\,\prime}\ .
\label{1.26}
\eeq
 This amplitude describes the
 diffractive transitions
 of
 the $\bar cc$ octet accompanied with change of the
 orbital momentum and
 parity of the $\bar cc$ pair.
 Since the color-exchange amplitude also has
 this property
 the product of these two amplitudes corresponds to
 creation
 of direct $J/\Psi$ ($\Psi'$...) on a nuclear
 target without radiation of any
 extra gluons.
 In this paper, however, we concentrate on the production of
 $\chi$s and neglect this effect, {\it i.e.} replace
 \beq
 \Sigma^{(8)}
 (\vec r_T,\vec
 r_T^{\,\prime},\alpha,\alpha^{\prime})
 \Rightarrow
 \sigma_3(\vec
 r_T, \alpha) + \sigma_3(\vec
 r_T^{\,\prime},\alpha^{\prime})\ .
\label{1.27}
\eeq
 
 If one uses the small-$r_T$ approximation (\ref{dipole}) also
 for
$\Sigma^{(tr)}$, then the cross section Eq.~({1.19}) can
 be represented in
the form,
 \beq
 \sigma(G\,A\to\chi\,X) = \int
d^2b\int\limits_{-\infty}^{\infty} dz\,\rho_A(b,z)\,
\sum\limits_{\lambda}\Bigl|A^{(\lambda)}(\vec b,z)
 \Bigr|^2\ ,
\label{1.28}
\eeq
 where
 \beq
 A^{(\lambda)}(\vec b,z)=\int\limits_0^1
 d\alpha\int
 d^2r_T\,
 \Phi_{\chi}^*(\vec r_T,\alpha)\,
 \hat A^{(\lambda)}(\vec b,z;\vec
 r_T,\alpha)\,
 \Phi_G(\vec r_T,\alpha)\ ,
\label{1.29}
\eeq
 and
 \beq
 \hat A^{(\lambda)}(\vec b,z;\vec r_T,\alpha) =
\exp\left[-{1\over2}\,\sigma_{\bar
 qq}(r_T)\,T_+(b,z)\right]\
 \vec
e_T^{\,(\lambda)}\cdot \vec d\
 \exp\left[-{1\over2}\,\sigma_3(r_T,\alpha)\,
T_-(b,z)\right]\ .
\label{1.30}
\eeq
 The vector $\vec d$ is defined in (\ref{1.5}).
 
 In Eq.~(\ref{1.30}) the first exponential factor
corresponds
to the nuclear attenuation of the $\chi$ produced in the
color-exchange rescattering of the projectile $\bar cc$.
 The second
exponential should be interpreted as nuclear
 shadowing for the diffractive
transition $G\to\bar cc$ which
 preserves the initial quantum numbers and
color (see
 in \cite{npz}). The factor
 $\vec e_T^{\,(\lambda)}\cdot \vec d$
corresponds to the
 amplitude of dipole radiation (or absorption) of a gluon
by
 the color-octet $\bar cc$ pair at the point $z$ of
the color-exchange
interaction.
 
\subsection{Medium high energies, $l_c\sim R_A$, but $l_f\gg
R_A$}\label{short-lc}
 
 Such an energy range is possible only for heavy
flavor production due to
 the relation $l_f\gg l_c$ \cite{kz91}. In this case
we can still treat the
 final state colorless $\bar cc$ pair propagating
through the nucleus as
 having a frozen transverse separation, while
corrections for finiteness of
 $l_c$ must be done. Correspondingly, the
operator $\hat A^{(\lambda)}(\vec
 b,z;\vec r_T,\alpha)$ has to be modified
and can be represented as a sum
 of two terms,
 \beq
 \hat
A^{(\lambda)}(\vec b,z;\vec r_T,\alpha) =
 \hat A_1^{(\lambda)}(\vec b,z;\vec
r_T,\alpha) +
 \hat A_2^{(\lambda)}(\vec b,z;\vec r_T,\alpha)\ ,
\label{2.1}
\eeq
 where
 \beq
 \hat A_1^{(\lambda)}(\vec b,z;\vec r_T,\alpha) =
 \exp\left[i\,q_L\,z - {1\over2}\,\sigma_{\bar qq}(r_T)\,
 T_+(b,z)\right]\,\vec e^{(\lambda)}\cdot\vec d\ ;
\label{2.2}
\eeq
 \beqn
 && \hat A_2^{(\lambda)}(\vec b,z;\vec r_T,\alpha) =
 -
 {1\over2}\,\exp\left[i\,q_L\,z - {1\over2}\,\sigma_{\bar
 qq}(r_T)\,
 T_+(b,z)\right]\,\vec e^{(\lambda)}_T\cdot\vec d
 \nonumber\\ &\times&
 \int\limits_{-\infty}^z dz_1\,
 \exp\left[i\,q_L\,z_1 -
 {1\over2}\,\sigma_{3}(r_T\alpha)\,
 T_-(b,z_1)\right]\,\sigma_3(\vec
 r_T,\alpha)\,
 \rho_A(b,z_1)\ ,
\label{2.3}
\eeqn
 where $q_L=1/l_c$ is the longitudinal momentum
 transfer to the
nucleus. This expression interpolates
 between the high-energy limit $l_c\gg
R_A$ ($q_L=0$) where
 it acquires the form (\ref{1.30}), and the limit of
$l_c$
 shorter than the mean nucleon separation when $\hat
A_2^{(\lambda)}\to 0$. Such a form of the amplitude
 Eqs.~(\ref{2.1}) -
(\ref{2.3}) was first derived in
 \cite{hkz} for inelastic diffractive
photoproduction of $J/\Psi$ off nuclei.
 
The term $A_1$ in Eq.~(\ref{2.1}) describes the
production of the colorless $\bar cc$ pair at ($b,z$) by the incident gluon.
This pair then attenuates escaping the nucleus, while the gluon 
has no initial state interaction (leading to production of a $\bar cc$ pair).
 
 The term $A_2$ in Eq.~(\ref{2.1}) describes the diffractive
 production of a $\bar cc$ pair by the gluon with the same
 quantum numbers (except the color dipole moment of the
$\bar
 cc$) at the point $z_1 < z$.  This color-octet $\bar cc$
 pair
propagates and attenuates between the points $z_1$ and
 $z$ with the cross
section $\sigma_3(r_T,\alpha)$.  Then at point $z$ it
 experiences a color exchange
interaction and
 produces a colorless $\bar cc$ pair with the
quantum numbers
 of a $\chi$.
 
 In fact, Eq.~(\ref{2.1}) breaks down towards
the low-energy limit at
 $\chi$ energies of a few tens of $GeV$ because $l_f$
becomes comparable
 with $R_A$ and one cannot neglect any more the
fluctuations of the
 transverse size of the $\bar cc$ pair during its
propagation through the
 nucleus. In this case one should apply the technique
of the light-cone Green function describing the 
propagation of $\bar qq$ pairs
\cite{krt1,z,kst1,kst2}.
 
\subsection{General case}\label{generalcase}

\begin{figure}[tbh]
 \includegraphics{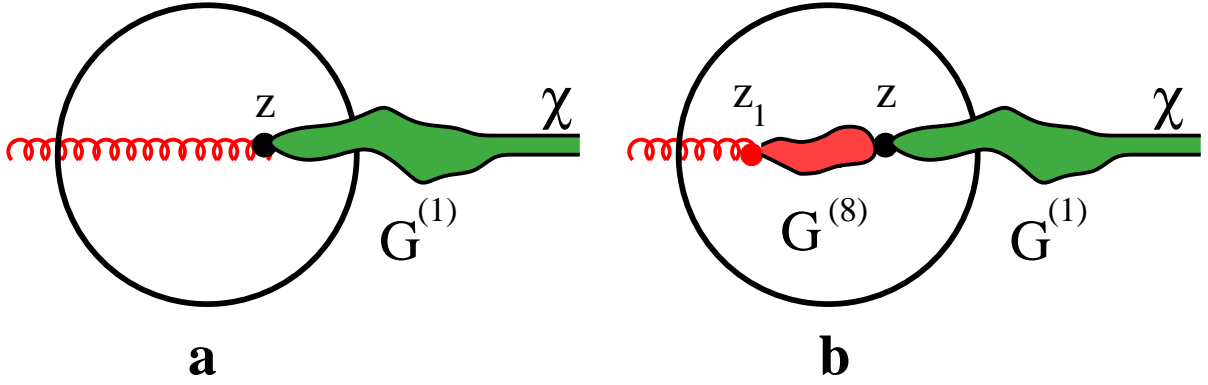}
\begin{center}
\vspace{6cm}
\parbox{13cm}
 {\caption[Delta]
 {\sl The incident gluon can either produce
the colorless
 $\bar cc$ pair with quantum numbers of $\chi$ at the point
$z$ ({\bf a}), or it produces diffractively
a color-octet $\bar cc$ with the quantum
 numbers of the gluon at the point $z_1$ which is then 
converted into a color singlet state at $z$ 
({\bf b}). Propagation of a
 color-singlet or octet $\bar cc$ is described by the
Green
 functions $G_{\bar cc}^{(1)}$ and $G_{\bar cc}^{(8)}$, respectively.}
\label{general-fig}}
\end{center}
 \end{figure}

 The transition between the limits of very short and very long coherence
lengths is performed using the prescription suggested in \cite{hkz} for
inelastic photoproduction of $J/\Psi$ off nuclei. The cross section of
 $\chi$
production off a nucleus can still be represented in the form
 (\ref{1.28}),
but the amplitude $A^{(\lambda)}$ is modified as,
 \beq
 A^{(\lambda)}(b,z)
=
 \int\limits_0^1 d\alpha \int d^2r_T\int
 d^2r_T^{\,\prime}\,
\Phi_{\chi}^*(\vec r_T,\alpha)\,
 \hat D^{(\lambda)}(\vec r_T,\vec
r_T^{\,\prime},
 \alpha;b,z)\, \Phi_G(\vec r_T^{\,\prime},\alpha)\ ,
\label{3.1}
 \eeq
 where $\hat D^{(\lambda)}(\vec r_T,\vec r_T^{\,\prime}, \alpha;b,z)$ is 
the amplitude of production of a colorless $\bar cc$ pair which
reaches a separation $\vec r_T$ outside the
 nucleus. It is produced at the
 point $(\vec b,z)$ by a
 color-octet $\bar cc$ with separation
 $r_T^{\,\prime}$.
 The amplitude consists of two terms,
 \beq
 \hat
 D^{(\lambda)}(\vec r_T,\vec
 r_T^{\,\prime},\alpha;b,z) =
 \hat
 D_1^{(\lambda)}(\vec r_T,\vec
 r_T^{\,\prime},\alpha;b,z) +
 \hat
 D_2^{(\lambda)}(\vec r_T,\vec
 r_T^{\,\prime},\alpha;b,z)\ .
\label{3.2}
 \eeq
 Here the first term reads,
 \beq
 \hat D_1^{(\lambda)}(\vec
 r_T,\vec
 r_T^{\,\prime},\alpha;b,z) =
 G_{\bar cc}^{(1)}(\vec r_T,z_+;\vec
 r_T^{\,\prime},z)\,
 \vec e^{\,(\lambda)}\cdot \vec d^{\,\prime}\,
 e^{iq_Lz}\ ,
\label{3.3}
 \eeq
 where $G_{\bar cc}^{(1)}(\vec r_T,z_+;\vec r_T^{\,\prime},z)$
 is the
color-singlet Green function describing evolution
 of a $\bar cc$ wave packet
with initial separation
 $\vec r_T^{\,\prime}$ at the point $z$ up to the
final
 separation $\vec r_T$ at $z_+\to\infty$. This term
 is illustrated in
Fig.~\ref{general-fig}a.
 
 There is also a possibility for the projectile gluon to
experience diffractive interaction with production of
 color-octet $\bar cc$
with the same quantum numbers of the
 gluon at the point $z_1$. This pair
propagates from
 the point $z_1$ to $z$ as is described by the corresponding
color-octet Green function $G_{\bar cc}^{(8)}$ and produces the final
colorless pair which propagation is described by the
 color-singlet Green
function, as is illustrated in
 Fig.~\ref{general-fig}b. The corresponding second
term in
 (\ref{3.2}) reads,
 \beqn
 \hat D_2^{(\lambda)}(\vec r_T,\vec
r_T^{\,\prime},\alpha;b,z) &=&
 - {1\over2}\,\int\limits_{-\infty}^z
dz_1\,d^2
 r_T^{\prime\prime}\,
 G_{\bar cc}^{(1)}(\vec
r_T,z_+;r_T^{\,\prime\prime},z)
 \nonumber\\ &\times&
 \vec
e^{(\lambda)}\cdot\vec d^{\,\prime\prime}\,
 G_{\bar cc}^{(8)}(\vec
r_T^{\,\prime\prime},z;
 \vec r_T^{\,\prime},z_1)\,
 e^{iq_Lz_1}\,
\sigma_3(\vec
 r_T^{\,\prime},\alpha)\,\rho_A(b,z_1)\ ,
\label{3.4}
 \eeqn
 
 The singlet, $G_{\bar cc}^{(1)}$, and octet, $G_{\bar cc}^{(8)}$,
Green
 functions describe the propagation  of                                    
 color-singlet  and octet $\bar cc$, respectively,
in the nuclear medium. They satisfy
 the Schr\"odinger equations,
\beq
 i\,\frac{d}{d\,z}\,
 G_{\bar cc}^{(k)}(\vec r_T,\vec
r_T^{\,\prime};z,z^\prime) =
 \left[\frac{m_c^2 -
 \Delta_{\vec
r_T}}{2\,E_G\,\alpha\,(1-\alpha)}\,
 + V^{(k)}(\vec r_T,\alpha)\right]\,
G_{\bar cc}^{(k)}(\vec r_T,\vec r_T^{\,\prime};z,z^\prime)\ ,
\label{3.5}
\eeq
 with $k=1,\ 8$ and boundary conditions
 \beq
 G_{\bar cc}^{(k)}(\vec
 r_T,\vec r_T^{\,\prime};z,z^\prime)
 \Bigl|_{z=z^\prime} =
 \delta(\vec
 r_T-\vec r_T^{\,\prime})\ .
\label{3.6}
\eeq
 
 The imaginary part of the LC potential $V^{(k)}$ is
 responsible for
the attenuation in nuclear matter,
 \beq
 {\rm Im}\,V^{(k)}(\vec r_T,\alpha)
=
 - {1\over2}\,\sigma^{(k)}(r_T,\alpha)\,
 \rho_A(b,z)\ ,
\label{3.7}
\eeq
 where
 \beqn
 \sigma^{(1)}(r_T,\alpha) &=& \sigma_{\bar qq}(r_T)\ ,
 \nonumber\\
 \sigma^{(8)}(r_T,\alpha) &=& \sigma_{3}(r_T,\alpha)\ .
\label{3.8}
\eeqn
 
 The real part of the LC potential $V^{(k)}(\vec r_T,\alpha)$
describes the interaction inside the $\bar cc$ system. For
 the singlet state
${\rm Re}\,V^{(1)}(\vec r_T,\alpha)$
 should be chosen to reproduce the
charmonium mass spectrum.
 With a realistic potential ({\it e.g.} see in
\cite{hikt})
 one can solve Eq.~(\ref{3.5}) only numerically.  Since this
paper is focused on the principal problems of understanding
 of the dynamics
of nuclear shadowing in charmonium
 production, we chose the oscillator form
of the potential
 \cite{kst2},
 \beq
 {\rm Re}\,V^{(1)}(\vec r_T,\alpha) =
\frac{a^4(\alpha)\,r_T^2}
 {2\,E_G\,\alpha\,(1-\alpha)}\ ,
\label{3.9}
 \eeq
 where
 \beqn
 a(\alpha)&=&2\,\sqrt{\alpha(1-\alpha)\,
 \mu\,\omega}\ ,
\label{3.10}\\
\mu&=&\frac{m_c}{2}\ ,\ \ \ \ \ \omega = 0.3\,GeV\ .
 \nonumber
 \eeqn
 The LC potential (\ref{3.9}) corresponds to a choice of
 a potential,
 \beq
U(\vec r)={1\over2}\,\mu\,\omega\,\vec r^2\ ,
\label{3.11}
 \eeq
 in the nonrelativistic Schr\"odinger equation,
 \beq
\left[-\frac{\Delta}{2\mu} +
 U(\vec r)\right]\,\Psi(\vec r) =
 E\,\Psi(\vec
r)\ ,
\label{3.12}
 \eeq
 which should describe the bound states of a colorless $\bar cc$ system.
Of course this is an approximation which we are enforced to do in order
to solve the evolution equation analytically.
 
To describe color-octet $\bar cc$ pairs we should fix the corresponding
potential at
 \beq
 {\rm Re}\,V^{(8)}(\vec r_T,\alpha)= 0\ ,
\label{3.13}
 \eeq
 in order to reproduce the gluon wave function Eq.~(\ref{1.02}).
 
\subsection{Numerical estimates}\label{numerics}

 In order to keep calculations simple we use the approximation
Eq.~(\ref{dipole}) for the dipole cross section which is
 reasonable for
small-size heavy quark systems.
 Then, taking into account Eqs.~(\ref{3.7}) -
(\ref{3.9})
 we arrive at the final expressions,
 \beq
 V^{(k)}(r_T,\alpha)
= {1\over2}\,
 \kappa^{(k)}\,r_T^2\ ,
\label{3.14}
 \eeq
 \beq
 \kappa^{(1)} =
 \frac{a^4(\alpha)}{\alpha(1-\alpha)\,E_G} -
 iC(s)\,\rho_A\ ,
\label{3.15}
 \eeq
 \beq
 \kappa^{(8)} = - iC(s)\,\rho_A\,
 \left\{{9\over8}\,\Bigl[\alpha^2+(1-\alpha)^2\Bigr]-
 {1\over8}\right\}\ .
\label{3.16}
 \eeq
 
 Making use of this approximation and assuming a constant nuclear
density
 $\rho_A(b,z)=\rho_A$ the Green functions can be obtained in an
analytical
 form,
 \beqn
 G_{\bar cc}^{(k)}(\vec r_T,\vec
r_T^{\,\prime};z_2,z_1) &=&
 \frac{b^{(k)}}{2\pi\,\sinh(\Omega^{(k)}\,\Delta
z)}
 \nonumber\\ &\times&
 \exp\left\{- \frac{b^{(k)}}{2}\,
\left[\frac{\vec r_T^{\,2} +\vec r_T^{\,\prime\,2}}
{\tanh(\Omega^{(k)}\,\Delta z)} -
 \frac{2\,\vec r_T\cdot\vec
r_T^{\,\prime}}
 {\sinh(\Omega^{(k)}\,\Delta z)}\right]\right\}\ ,
\label{3.17}
 \eeqn
 where
 \beqn
 b^{(k)} &=& \sqrt{\kappa^{(k)}\,
 E_G\,\alpha(1-\alpha)}\ ,\nonumber\\
 \Omega^{(k)} &=&
 \frac{b^{(k)}}{E_G\,\alpha(1-\alpha)}\ ,
 \nonumber\\
 \Delta z &=& z_2-z_1\
 .
 \nonumber
 \eeqn
 
 With the oscillator potential we have chosen the
 wave function
 of $\chi$ has a simple form \cite{kst2},
 \beq
 \Phi_{\chi}(\vec r_T,\alpha) \propto
 (\vec e_+\cdot\vec
 r_T)\,\exp\left(-{1\over2}\,
 a^2(\alpha)\,r_T^2\right)\ ,
\label{3.18}
 \eeq
 which allows to perform analytical integrations over $\vec r_T$ and
 $\vec r_T^{\,\prime}$ in Eq.~(\ref{3.1}) for the amplitude,
 \beq
 A^{(\lambda)}(b,z) = N(\vec e^+\Bigr.^*\cdot\vec
 e^{\lambda})\,\int\limits_0^1 d\alpha\,
 \left[U(\vec b,z,\alpha)\,e^{iq_Lz}
 -
 \frac{c^{(8)}}{2}\int\limits_{-R(\vec b)}^z
 dz_1\,\rho_A(\vec b,z)\,
 e^{iq_Lz_1}\,W(\vec b,z,z_1,\alpha)\right]\ .
\label{3.19}
 \eeq
 Here
 \beq
 U(b,z,\alpha) =
 \frac{F_1(\lambda_1)}
 {\Bigl[b^{(1)}\,\sinh(\phi_1) +
 a^2(\alpha)\,\cosh(\phi_1)\Bigr]^2}\ ;
\label{3.20}
 \eeq
 
 \beqn
 W(b,z,z_1,\alpha) &=&
\frac{\sinh(\phi_2)\,F_1(\lambda_2)}
{b^{(8)}\,\Bigl[b^{(8)}\,\sinh(\phi_2)+
 g_2\,\cosh(\phi_2)\Bigr]^2}
\nonumber\\
 &+& \frac{F_2(\lambda_2)}
{\Bigl[b^{(8)}\,\sinh(\phi_2)+g_2\,\cosh(\phi_2)\Bigr]^3}\ ,
\label{3.22}
 \eeqn
 where
 \beqn
 F_1(\lambda_1) &=& (1+\lambda_1)\,e^{\lambda_1}\,
 E_1(\lambda_1)\,-\,1\ ,
 \nonumber\\
 F_2(\lambda_2) &=&
 (\lambda_2^2+4\,\lambda_2+2)\,
 e^{\lambda_2}\,E_1(\lambda_2) -
 \lambda_2 -
 3\ ,
\label{3.23}
 \eeqn
 and $E_1(x)=\int_1^\infty dt\,e^{-xt}/t$
 is the integral exponential
 function.
 Further notations are
 \beqn
 \phi_1 &=& \Omega^{(1)}\,\Delta
 z_1\ ,
mueller \nonumber\\
 \Delta z_1 &=& \sqrt{R_A^2-b^2} - z\ ,
 \nonumber\\
 \phi_2 &=& \Omega^{(8)}\,\Delta z_2\ ,
 \nonumber\\
 \Delta z_2 &=& z-z_1\
 ,
 \nonumber
 \eeqn
 \beqn
 \lambda_1 &=& \frac{m_c^2\,
 \Bigl[a^2(\alpha)\,\sinh(\phi_1) +
 b^{(1)}\,\cosh(\phi_1)\Bigr]}{2\,b^{(1)}\,
 \Bigl[a^2(\alpha)\,\cosh(\phi_1)
 +
 b^{(1)}\,\sinh(\phi_1)\Bigr]}\ ,
 \nonumber\\
 \lambda_2 &=&
 \frac{m_c^2\,
 \Bigl[g\,\sinh(\phi_2) +
 b^{(8)}\,\cosh(\phi_2)\Bigr]}{2\,b^{(8)}\,
 \Bigl[g\,\cosh(\phi_2) +
 b^{(8)}\,\sinh(\phi_2)\Bigr]}\ ,
 \nonumber\\
 g &=&
 \frac{m_c^2}{2\,\lambda_1}\ ,
 \nonumber\\
 N &=& \frac{1}{2\,\pi}\,
 \sqrt{\frac{\alpha_s\,C\,(\mu\omega)^{5/2}}{m_c}}\ .
 \nonumber
 \eeqn
 
\begin{figure}[tbh]
 \includegraphics{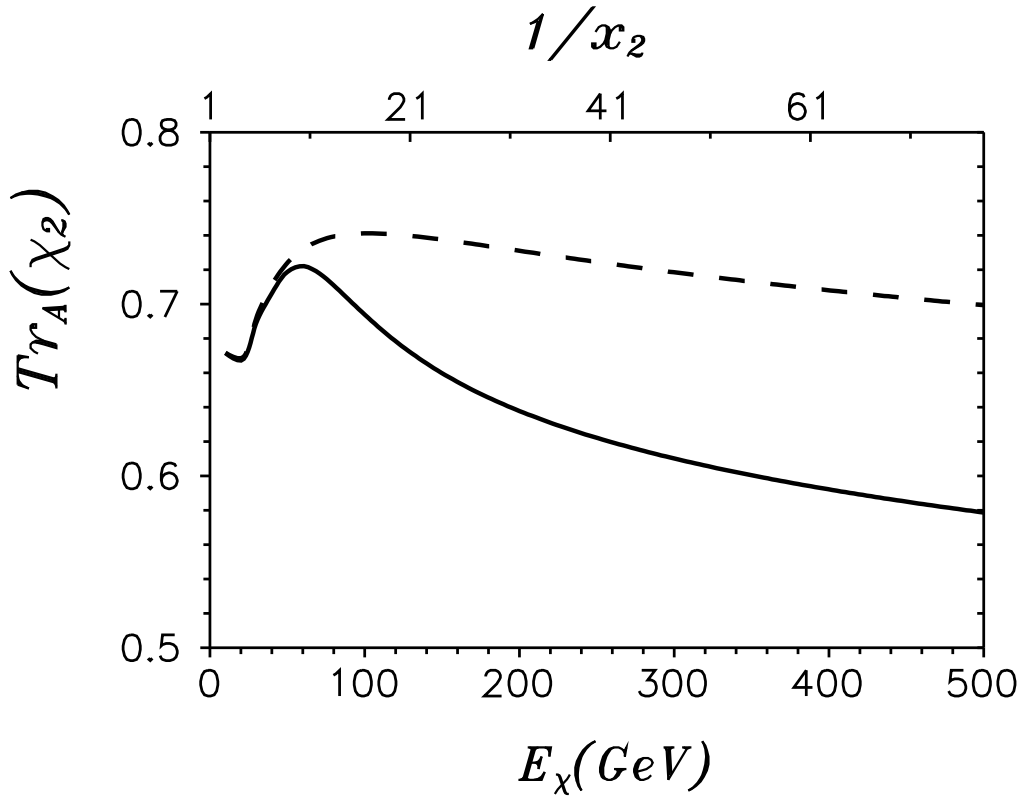}
\begin{center}
\vspace{11cm}
\parbox{13cm}
 {\caption[Delta]
 {\sl Nuclear transparency for $\chi$ production off lead as function
of energy of the charmonium, or $x_2$ (the upper scale).  The solid curve
includes both effects of coherence and formation, while the dashed curve
corresponds to $l_c=0$. Since transparency scales in $x_2$ according to
(\ref{scaling}), values of $x_2$ are shown on the top axis.}
 \label{e-dep}}
\end{center}
 \end{figure}

 We define the nuclear transparency for $\chi$ production as
 \beq
 Tr_A(\chi)=\frac{\sigma(G\,A \to \chi\,X)}
 {A\,\sigma(G\,N \to
 \chi\,X)}\ .
\label{transparency}
 \eeq
 It depends only on the $\chi$ or projectile gluon energy. We plot our
 predictions for lead in Fig.~\ref{e-dep}.
 Transparency rises at low energy since the formation length
 increases
 and the effective absorption cross section becomes smaller. This
 behavior, assuming $l_c=0$, is shown by dashed curve. However, at higher
 energies the coherence length is switched on and shadowing adds to
 absorption in accordance to Eq.~(\ref{1.30}). As a result, transparency
 decreases, as is shown by the solid curve. On top of that, the 
energy dependence of the dipole cross section makes those both curves 
for $Tr_a(E_{\chi})$ fall even faster.
 
Apparently, the nuclear  transparency depends only on the $\chi$ energy, 
rather than the incident energy or $x_1$. It is interesting that it lead 
to the $x_2$ scaling. Indeed, the $\chi$ energy
 \beq
E_{\chi}=\frac{M_{\chi}^2}{2\,m_N\,x_2}\ ,
\label{scaling}
 \eeq
depends only on $x_2$. We show the $x_2$ scale in Fig.~\ref{e-dep}
(top) along with energy dependence.

\begin{figure}[tbh]
 \includegraphics{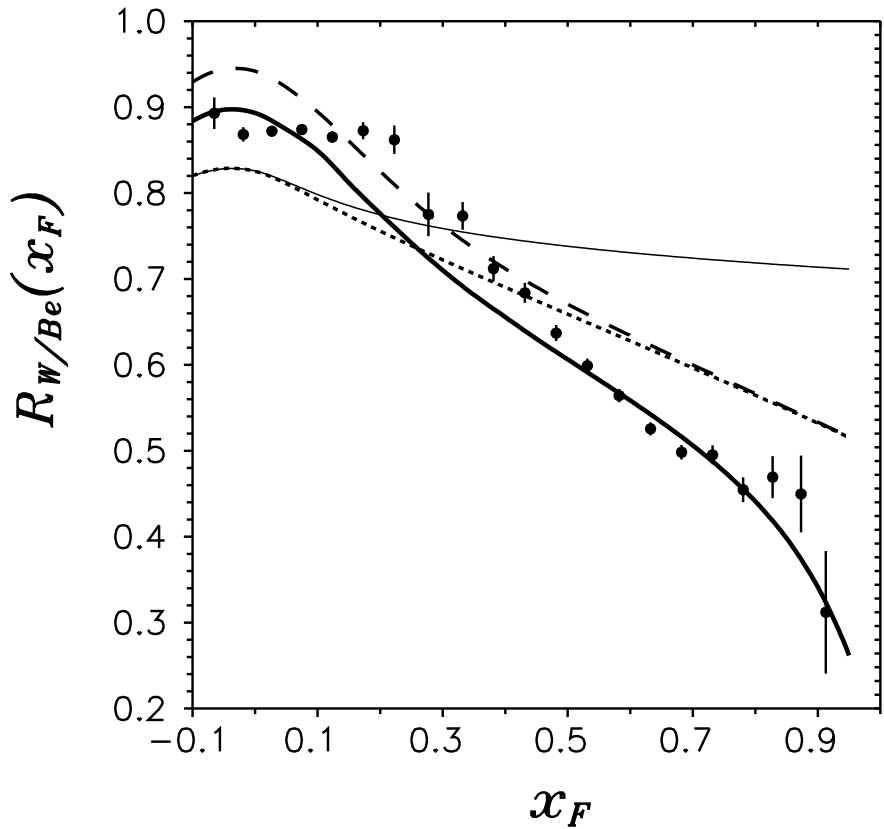}
\begin{center}
\vspace{11cm}
\parbox{13cm}
 {\caption[Delta]
 {\sl Tungsten to beryllium cross section ratio as function of Feynman
$x_F$ for $J/\Psi$ production at proton energy $800\,GeV$. The thin
solid curve represents contribution of initial state quark shadowing and
finals state $\bar cc$ attenuation for $\chi$ production. The dotted
curve includes also gluon shadowing. The dashed curve is corrected for
gluon enhancement at large $x_2$ (small $x_F$) using the prescription
from \cite{ekr}.  The final solid curve is also corrected for energy
loss and for $\chi\to J/\Psi\gamma$ decay.  Experimental points are from
the E866 experiment \cite{e866}.}
 \label{xf-dep}}
\end{center}
 \end{figure}

We also compare in Fig.~\ref{xf-dep} the contribution of quark shadowing 
and absorption (thin solid curve) with the 
nuclear suppression observed at $800\,GeV$ \cite{e866}.
Since data are  for $W/Be$  ratio, and our constant density approximation 
should not be applied to beryllium, we assume for simplicity that all 
$pA$ cross sections including $pN$ obey the $A^{\alpha(x_F)}$.
We see  that the calculated contribution has quite a different shape
from what is suggested by the data. It also leaves plenty of room for 
complementary mechanisms of suppression at large $x_F$.

\section{Gluon shadowing}\label{gluons}

Previously we considered only the lowest $|\bar cc\ra$ fluctuation
of the gluon what is apparently an approximation. The higher Fock components
containing gluons should be also included. In fact they are already incorporated
in the phenomenological dipole cross section we use, and give rise to the energy
dependence of $\sigma^N_{\bar qq}$. However, they are still excluded from nuclear
effects. Indeed, although we eikonalize the energy dependent dipole cross section
the higher Fock components do not participate in that procedure, but 
they have to be eikonalized as well. This corrections, as is demonstrated below, 
correspond to suppression of gluon density in nuclei at small $x$.

The gluon density at small $x$ in nuclei is known to be shadowed,
i.e. reduced compared to a free nucleons. The partonic
 interpretation of this phenomenon looks very different dependent on
 reference frame. In the infinite
momentum frame, as was first
 suggested by Kancheli \cite{kancheli}, the
partonic clouds of
 nucleons are squeezed by the Lorentz transformation less
at small than
 at large $x$. Therefore, while these clouds are well separated in
longitudinal
 direction at large $x$, they overlap and can fuse at small $x$,
resulting
 in a diminished parton density \cite{kancheli,glr}.
 
 Different
observables can probe this effect. Nuclear shadowing of the
 DIS inclusive
cross section or Drell-Yan process demonstrate a reduction
 of the sea quark
density at small $x$. Charmonium or open charm
 production is usually
considered as a probe for gluon distribution.
 
 Although observables are
Lorentz invariant, partonic interpretation is
 not, and the mechanism of
shadowing looks quite different in the rest
 frame of the nucleus where it
should be treated as Gribov's inelastic
 shadowing.  This approach seems to go
better along with our intuition,
 besides, the interference or coherence
length effects governing
 shadowing are under a better control.  One can even
calculate
 shadowing in this reference frame in a parameter free way (see in
\cite{krt2,kst1,kst2}) employing the well developed phenomenology of
 color
dipole representation suggested in \cite{zkl}.  On the other
 hand, within the
parton model one can only calculate the $Q^2$
 evolution of shadowing which is
quite a weak effect.  The main
 contribution to shadowing originates from the
fitted to data input.
 
 In the color dipole representation nuclear shadowing can be
calculated via simple eikonalization of the elastic amplitude for each
Fock component of the projectile light-cone wave function which are
the eigenstates of interaction \cite{zkl}. Different Fock components
represent shadowing of different species of partons. The $|\bar qq\ra$
component in DIS or $|q\gamma^*\ra$ in Drell-Yan reaction should be
used to calculate shadowing of sea quarks. The same components
including also one or more gluons lead to gluon shadowing
\cite{mueller,kst2}.

In the color dipole approach one can explicitly see deviations from
QCD factorization, i.e. dependence of the measured parton
distribution on the process measuring it. For example, the coherence
length and nuclear shadowing in Drell-Yan process vanish at minimal
$x_2$ (at fixed energy) \cite{eloss,eloss1}, while the factorization predicts
maximal shadowing. Here we present even more striking deviation from
factorization, namely, gluon shadowing for charmonium production turns
out to be dramatically enhanced compared to DIS.

\subsection{LC dipole representation for reaction 
\boldmath$G\,N \to \chi\,G\,X$}\label{lc-wf}
 
\begin{figure}[tbh]
 \includegraphics{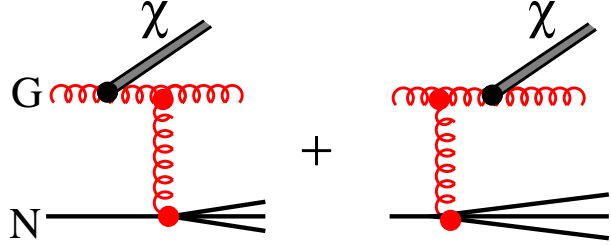}
\begin{center}
\vspace{3.5cm}
\parbox{13cm}
 {\caption[Delta]
 {\sl The dominant Feynman diagrams contributing to $\chi$ 
production (see \ref{diagrams}).}
 \label{dy}} 
\end{center}
 \end{figure}

 In the case of charmonium production different Fock components of the
projectile gluon, $|(\bar cc)_1nG\ra$ containing a colorless $\bar cc$
pair and $n$ gluons (n=0,1...) build up the cross section of charmonium 
production which steeply rises with energy
(see in \cite{hikt}). The cross 
section is expected to factorize in impact parameter representation in
analogy to the DIS and Drell Yan reaction. This representation has the
essential advantage in that nuclear effects can be easily calculated 
\cite{hir,kst1}. Feynman diagrams corresponding $\chi$ production associated
with gluon radiation are depicted in Fig.~\ref{graphs} of the Appendix~D. 
We treat the interaction of heavy quarks perturbatively in the lowest order
approximation, while the interaction with the nucleon is soft and
expressed in terms of the gluon distribution. The calculations
performed in \ref{diagrams} are substantially simplified if the
radiated gluon takes a vanishing fraction $\alpha_3$ of the total
light-cone momentum and the heavy quarkonium can be treated as a
nonrelativistic system. In this case the amplitude of $\chi\, G$
production has the simple form Eq.~(\ref{d25}) which corresponds to
the ``Drell-Yan'' mechanism of $\chi$ production illustrated in 
Fig.~\ref{dy}.
 Correspondingly, the cross section of $\chi$ production has the 
familiar factorized form similar to the Drell-Yan reaction 
\cite{hir,bhq,kst1},
 \beq
\alpha_3\,\frac{d\,\sigma(GN\to\chi G X)}
{d\,\alpha_3} = 
\int d^2s\, \left|\Psi_{G\chi}(s,\alpha_3)\right|^2\,
\sigma_{GG}\left[(1-\alpha_3)\vec s,x_2/\alpha_3\right]\ ,
\label{3.2a}
 \eeq
 where $\sigma_{GG}(r,x)=9/4\,\sigma_{\bar qq}(r,x)$ [see
Eq.~(\ref{1.3a})] is the cross section of interaction of a $G\,G$
dipole with a nucleon. $\Psi(s,\alpha_3)$ is the effective
distribution amplitude for the $\chi-G$ fluctuation of a gluon, which
is the analog to the $\gamma^*\,q$ fluctuation of a quark,
 \beqn
\Psi_{G\chi}(s,\alpha_3) &=& \sum\limits_{\bar\mu\mu}\,
\int d^2r\,d\alpha\, 
\Phi^{\bar\mu\mu}_{\chi}(\vec r,\alpha)\,
\Phi^{\bar\mu\mu}_{G}(\vec r,\alpha) 
\nonumber\\ &\times&
\left[\Phi_{c G}\left(\vec s + 
\frac{\vec r}{2},\frac{\alpha_3}{\alpha}\right)
- \Phi_{c G}\left(\vec s - 
\frac{\vec r}{2},\frac{\alpha_3}
{1-\alpha}\right)\right]\ .  
\label{3.3a}
 \eeqn 
 The notations of \ref{diagrams} are used here,
$\Phi^{\bar\mu\mu}_{\chi}(\vec r,\alpha)$ and $\Phi^{\bar\mu\mu}_{G}(\vec
r,\alpha)$ are the $\bar qq$ LC wave functions of the $\chi$ and gluon,
respectively, which depend on transverse separation $\vec r$ and relative
sharing $\alpha$ by the $\bar qq$ of the total LC momentum. $\Phi_{\chi
G}(\vec s,\alpha)$ is the LC wave function of a quark-gluon Fock
component of a quark given by Eq.~(\ref{d21}).

\subsection{Gluon shadowing for \boldmath$\chi$ production off 
nuclei}\label{gluon-shadowing}

The gluon density in nuclei is known to be modified, shadowed at small
Bjorken $x$. Correspondingly, production of $\chi$ treated as gluon-gluon
fusion must be additionally suppressed. In the infinite momentum frame of
the nucleus gluon shadowing should be treated as a $G\,G \to G$ fusion
which diminishes the amount of gluons \cite{kancheli,glr}. In the rest frame
of the nucleus it looks very different as Gribov's inelastic shadowing
\cite{gribov} related to diffractive gluon radiation. This frame seems to
be more convenient way to calculate gluon shadowing, since techniques are
better developed, and we use it in what follows.

The process $G\,N \to \chi\,X$ considered in the previous section 
includes by default radiation of any number of gluons which give rise to
the energy dependence of the dipole cross section in (\ref{1.3a})
 
Extending the analogy between the reactions of $\chi\,G$ production by an
incident gluon and heavy photon radiation by a quark to the case of
nuclear target one can write an expression for the cross section
of reaction $G\,A\to\chi\,G\,X$ in two limiting cases:

{\bf (i)} the production occurs nearly instantaneously over a
longitudinal distance which is much shorter than the mean free path of
the $\chi\,G$ pair in nuclear matter. In this case the cross sections on
a nuclear and nucleon targets differ by a factor $A$ independently of the
dynamics of $\chi\,G$ production.

{\bf (ii)} The lifetime of the $\chi\,G$ fluctuation,
 \beq
t_c = \frac{2\,E_G}{M^2_{\chi G}}\ ,
\label{3.4a}
 \eeq
substantially exceeds the nuclear size. It is straightforward to 
replace the dipole cross section on a nucleon by a nuclear one 
\cite{hir,kst1},
then Eq.~(\ref{3.2a}) is modified to
 \beq
\frac{d\,\sigma(GA\to\chi G X)}
{d\,(\ln\alpha_3)} = 
2\int d^2b\,d^2s\, \left|\Psi_{G\chi}(\vec s,\alpha_3)\right|^2\,
\left\{1 - \exp\left[-{1\over2}\,\sigma^N_{GG}(\vec s,x_2/\alpha_3)\,
T_A(b)\right]\right\}\ .
\label{3.5a}
 \eeq
 In order to single out the net gluon shadowing we exclude here the size
of the $\bar cc$ pair assuming that the cross section responsible for
shadowing depends only on the transverse separation $\vec s$.

A general solution valid for any value of $t_c$ is more complicated
and must interpolate between the above limiting situations.\\
{\bf (iii)} In this case one can use the methods of the
Landau-Pomeranchuk-Migdal (LPM) theory for photon bremsstrahlung in a
medium generalized for targets of finite thickness in \cite{z,kst1}.
The general expressions for the cross section which reproduces the 
limiting cases $t_c\to0$ ({\bf i}) and $t_c\to\infty$ ({\bf ii})
reads,
 \beqn
\frac{d\,\sigma(GA\to\chi G X)}
{d\,(\ln\alpha_3)} \,
&=& \int d^2b\,\left\{\int\limits_{-\infty}^{\infty}
dz\,\rho_A(b,z)\,\int d^2s\,
\left|\Psi_{G\chi}(s,\alpha_3)\right|^2\right.
\sigma_{GG}\left[(1-\alpha_3)\vec s,x_2/\alpha_3\right]
\nonumber\\ 
&-& \left. {1\over2}\,{\rm Re}
\int\limits_{-\infty}^{\infty}dz_2\,
\rho_A(b,z_2)
\int\limits_{-\infty}^{z_2}dz_1\,
\rho_A(b,z_1)\,
\widetilde\Sigma(z_2,z_1)\,
e^{iq_L(z_2-z_1)}\right\}\ ,
\label{3.6a}
 \eeqn
 where
 \beq
\widetilde\Sigma(z_2,z_1) = 
\int d^2s_1\,d^2s_2\,
\Psi_{G\chi}^*(\vec s_2,\alpha_3)\,
\sigma^N_{GG}(s_2,x_2/\alpha_3)\,
G(\vec s_2,z_2;\vec s_1,z_1)\,
\sigma^N_{GG}(s_1,x_2/\alpha_3)\,
\Psi_{G\chi}(\vec s_1,\alpha_3)\ .
\label{3.7a}
 \eeq
 Here the Green function $G(\vec s_2,z_2;\vec s_1,z_1)$ describes
propagation of the $\chi\,G$ pair which starts with transverse separation
$\vec s_1$ at the points with longitudinal coordinates $z_1$ and ands up
at $z_2$ having separation $\vec s_1$. It satisfies the Schr\"odinger
type equations,
 \beq
i\,\frac{\partial G(\vec s_2,z_2;\vec s_1,z_1)}
{\partial z_2} = \left[-\frac{\Delta_{s_2}}
{2E_G\alpha_3(1-\alpha_3)} - i\,\sigma^N_{GG}(s_2)\,
\rho_A(B,z_2)\right]\,
G(\vec s_2,z_2;\vec s_1,z_1)\ ;
\label{3.8a}
 \eeq
 \beq
i\,\frac{\partial G(\vec s_2,z_2;\vec s_1,z_1)}
{\partial z_1} = - \left[\frac{\Delta_{s_1}}
{2E_G\alpha_3(1-\alpha_3)} - i\,\sigma^N_{GG}(s_1)\,
\rho_A(B,z_1)\right]\,
G(\vec s_2,z_2;\vec s_1,z_1)\ .
\label{3.9a}
 \eeq
 In the limit $z_2-z_1 \to 0$ it should satisfy the condition,
 \beq
G(\vec s_2,z_2;\vec s_1,z_1)\Bigr|_{z_2=z_1} = 
\delta(\vec s_2-\vec s_1)\ .
\label{3.10a}
 \eeq

A full calculation of the cross section of associated $\chi\,G$
production off nuclei Eq.~(\ref{3.6a}) employing the solutions of
equations (\ref{3.7a}) - (\ref{3.10a}) with realistic shapes of the dipole
cross section and nuclear density can be done only numerically, and is
still a challenge for computing. However, the problem can be essentially
simplified if the following approximations are done,
 \beqn
\rho_A(b,z) &=& \Theta\left(R_A-\sqrt{b^2+z^2}\right)\ ;
\nonumber\\
\sigma_{GG}(s,x) &\approx& C_{G}(x)\,s^2\ .
\label{3.11a}
 \eeqn
 The former is rather realistic for heavy nuclei, while the latter needs
special care to be adjusted to realistic calculations, since the value of
factor $C_{G}(x)$ correlates with the mean separation $\la s\ra$ which
depends on the process.

Calculations of the nuclear cross section with the realistic dipole cross
section which levels off at large separations is complicated only for
the transition region of $l_c\sim R_A$. In the limit of $l_c\to \infty$ one
can easily perform calculations with any form of the dipole cross section
and adjust the $C_G(x)$ to reproduce the cross section of reaction
$G\,A\to\chi\,G\,X$.  Such a prescription guarantees the correct endpoint
behavior at $t_c\gg R_A$ and $t_c\to 0$, then we expect the transition
region should not be very wrong either.

\begin{figure}[tbh]
\includegraphics{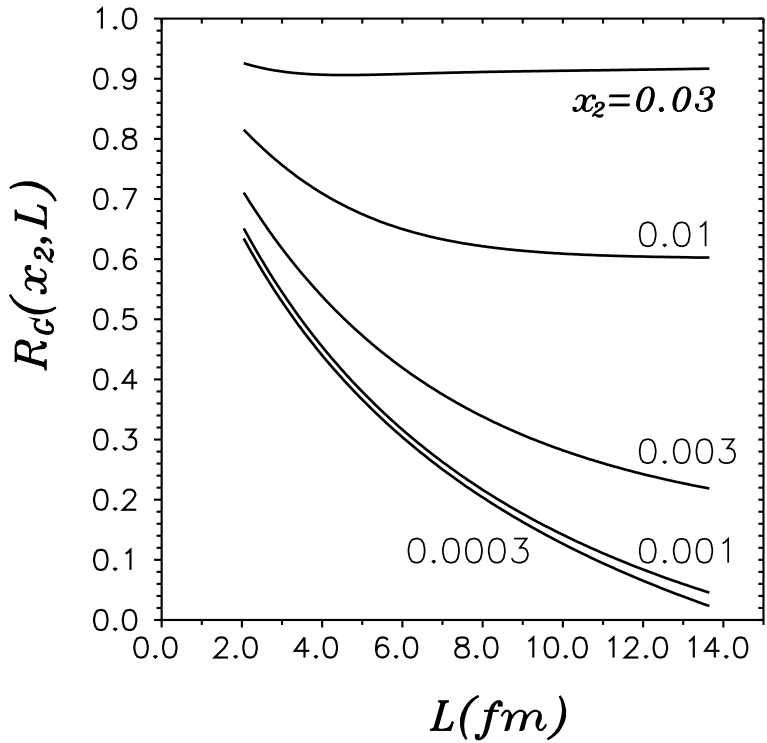}
\begin{center}
\vspace{10cm}
\parbox{13cm}
{\caption[Delta]
{\sl Gluon suppression as function of thickness of nuclear matter
with constant density $\rho_A=0.16\,fm^{-3}$.}
\label{glue-shad}}
\end{center}
 \end{figure}

To single out the correction for gluon shadowing one should compare the 
cross section Eq.~(\ref{3.6a}) with the impulse approximation term in 
which absorption is suppressed,
 \beq
R_G(x_2)=
\frac{G_A(x_2)}{A\,G_N(x_2)} = 
1 - \frac{1}{A\,\sigma(GN\to\chi X)}\,
\int\limits_{x_2}^{\alpha_{max}}
d\alpha_3\,\frac{d\sigma(GA\to\chi GX)}
{d\alpha_3}\ .
\label{3.12a}
 \eeq
 For further calculations and many other applications one needs to know
gluon shadowing as function of impact parameter which is calculated as 
follows,
 \beq
R_G(x_2,b)=
\frac{G_A(x_2,b)}{T_A(b)\,G_N(x_2)} =   
1 - \frac{1}{T_A(b)\,\sigma(GN\to\chi X)}\,
\int\limits_{x_2}^{\alpha_{max}}d\alpha_3\,
\frac{d\sigma(GA\to\chi GX)}                                        
{d^2b\,d\alpha_3}\ ,
\label{3.15a}
 \eeq
 where $d\sigma(GA\to\chi GX)/d^2b\,d\alpha_3$ is the function integrated
over $d^2b$ in (\ref{3.6a}). The results of calculations for the 
$b$-dependence of gluon shadowing (\ref{3.15a}) are depicted in
Fig.~\ref{glue-shad} for different values of $x_2$ as function of thickness
of nuclear matter, $L=\sqrt{R_A^2-b^2}$.
 The results confirm the obvious expectation that shadowing increases for
smaller $x_2$ and for longer path in nuclear matter. One can see that for
given thickness shadowing tends to saturate down to small $x_2$, what
might be a result of one gluon approximation. Higher Fock components with
larger number of gluons are switched on at very small $x_2$. At the same
time, shadowing saturates at large lengths what one should have also
expected as a manifestation of gluon saturation. Note that at large
$x_2=0.03$ shadowing is even getting weaker at longer $L$. This is easy
to understand, in the case of weak shadowing one can drop off the
multiple scattering terms higher than two-fold one. Then the shadowing
correction is controlled by the longitudinal formfactor of the nucleus
which decreases with $L$ (it is obvious for the Gaussian shape of the
nuclear density, but is also true for the realistic Woods-Saxon
distribution).

The next problem we face is how to correct our previous calculations for 
the calculated gluon shadowing. The standard parton model prescription is 
to multiply the cross section of charmonium production by $R_G(x_2)$.
This may be correct only if the process is so hard that no nuclear 
effects except gluon shadowing exists. For instance, this is the case in 
DIS for highly virtual longitudinally polarized photons. However, as we 
calculated and Fig.~\ref{e-dep} demonstrates, multiple interaction of the 
$\bar cc$ pair cannot be ignored, and the parton model prescription 
is invalid. To get a correct result one should perform a full calculation 
of nuclear effects which involve all Fock components. This is still a 
challenge, meanwhile one should look for a reasonable approximation.

\begin{figure}[tbh]
\includegraphics{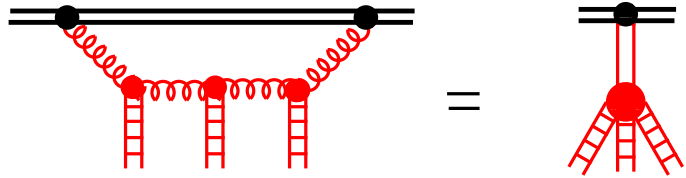}
\begin{center}
\vspace{2.8cm}
\parbox{13cm}
{\caption[Delta]
{\sl Multiple rescattering of a gluonic component of a hadron in the 
target rest frame. In $t$-channel it is interpreted as multi-Pomeron 
fusion $n\Pom \to \Pom$.}
\label{fusion}}
\end{center}
 \end{figure}

\begin{figure}[tbh]
\includegraphics{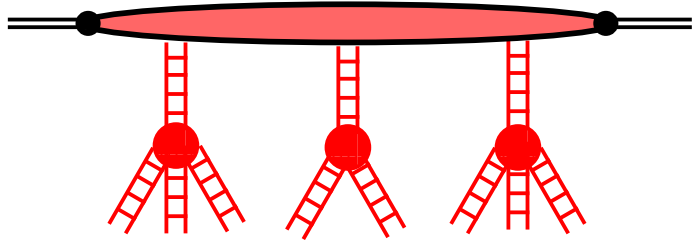}
\begin{center}
\vspace{4cm}
\parbox{13cm}
{\caption[Delta]
{\sl Multiple rescattering of a projectile hadron (e.g. $\bar cc$)
via multi-Pomeron interactions diminished by $n\Pom \to \Pom$
fusion.}
\label{pomerons}}
\end{center}
 \end{figure}

Our approximation of the lowest Fock state containing only one gluon 
corresponds
in term of Regge theory to Pomeron fusion $n\Pom \to \Pom$ (n=2,3...),
while transitions $n\Pom \to m\Pom$ ($m\geq2$) are missed. This is 
explained pictorially in Fig.~\ref{fusion}.
 Although inclusion of the multi-gluon Fock states is problematic, there
is a simple and intuitive way to include an essential part of missed
corrections. Intuitively one can say that the projectile high energy
partons experience multiple interactions not with target nucleons, but
with gluons at small $x_2$. Eikonal (Glauber)  approximation assumes that
the number of gluons is proportional to the number of nucleons. This is
why the eikonal exponent contains a product of the interaction cross
section and nuclear thickness, $\sigma\,T_A(B)$. It is not true any more
if gluon fusion is at work. The gluon density decreases (the more, the
larger the $T_A$ is) and one can easily take it into account
renormalizing the nuclear density,
 \beq
\rho_A(b,z) \Rightarrow \tilde\rho_A(b,z) = 
R_G(x_2,b)\,\rho_A(b,z)\ .
\label{3.20a}
 \eeq
 In this way we do account for the missed $n\Pom \to m\Pom$ transitions.
Indeed, $m$-fold interaction associated with $m\Pom$ exchange is not
are not in sequence, this would lead to the famous AFS (Amati-Fubini-Stangelini)
cancelation. The $m$ Pomerons correspond to simultaneous development of
$m$ parton ladders. In our approximation each of this Pomeron ladder is
suppressed by fusion precess which is exactly $n^\prime\Pom \to \Pom$ in
this case. By eikonalizing [see e.g.Eq.~(\ref{1.30})] the result of 
fusion we correctly involve the higher orders of $n^\prime\Pom \to 
1\Pom$, as is illustrated in Fig.~\ref{pomerons}.

Thus, we correct our results for for gluon shadowing renormalizing the
nuclear density by means of (\ref{3.20a}). It must be done in the
imaginary part of the light-cone potential (\ref{3.7}) in the evolution
equation Eq.~(\ref{3.5}) and also in Eqs.~(\ref{1.28}) and (\ref{3.4}).
The numerical results are depicted by dotted curve in Fig.~\ref{xf-dep}.
Apparently, gluon shadowing is stronger at small $x_2$, i.e. at large 
$x_F$.

\subsection{Antishadowing of gluons}\label{antishadowing}

Nuclear modification of the gluon distribution is poorly known.
There is still no experimental evidence for that. Nevertheless,
the expectation of gluon shadowing at small $x$ is very solid, and only 
its amount might be disputable. At the same time,
some indications exist that gluons may be enhanced in nuclei 
at medium small $x_2\sim 0.1$. The magnitude of gluon antishadowing
has been estimated in \cite{fs} assuming that the total fraction of 
momentum carried by gluons is the same in nuclei and free nucleons
(there is an experimental support for it). Such a momentum conservation
sum rule leads to a qluon enhancement at medium $x$, since gluons
are suppressed in nuclei at small $x$. The effect, up  to $\sim 20\%$
antishadowing in heavy nuclei at $x\approx 0.1$, found in \cite{fs} 
is rather large, but it is a result of very strong shadowing which
we believe has been grossly overestimated (see discussion in
\cite{kst2}).

Fit to DIS data based on evolution equations performed in
\cite{ekr} also provided an evidence for rather strong antishadowing
effect at $x\approx 0.1$. However, the fit 
employed an ad hoc assumption that gluons 
are shadowed at the low scale $Q_0^2$  
exactly as $F_2(x,Q^2)$ what might be true only by accident. Besides, in 
the $x$ distribution of antishadowing was shaped ad hoc too.

A similar magnitude of antishadowing has been found in the analysis 
\cite{gp} of data on $Q^2$ dependence of nuclear to nucleon ratio of the
structure functions, $F^A_2(x,Q^2)/F^N_2(x,Q^2)$. However it was based on the 
leading order QCD approximation which is not well justified at these values 
of $Q^2$. 

Although neither of  these results seem to be reliable, 
similarity of the scale of the predicted effect
looks convincing, and we included the antishadowing of gluons 
in our calculations. We use the shape of $x_2$ dependence and magnitude of 
gluon enhancement from \cite{ekr}.

\section{Energy loss corrections}\label{e-loss}\label{eloss}

 The mechanism of $J/\Psi$ suppression at large $x_F$ due to initial
state energy loss of the projectile partons in nuclei was first
 suggested back in
1984 \cite{kn1}. As soon as the incoming hadron
 interacts inelastically in
the nucleus, the projectile partons
 (only those which are sufficiently soft
to be resolved by the soft
 interaction) lose coherence and start losing
energy for
 hadronization. If the coherence time of charmonium production is
shorter than the mean internucleon separation in a nucleus, as it
 was assumed
in \cite{kn1}, the projectile parton (gluon)
 responsible for charmonium
production arrives at the production
 point with a diminished energy
$E_G-\Delta E_G$. In order to
 produce a charmonium with the same energy
(fixed by the measurement)
 as on a proton target the projectile gluon must
have an excess in
 the initial momentum which leads to a suppression
 \beq
R_G(x_1,\Delta x_1) =
 \frac{g^h(x_1+\Delta x_1)}{g^h(x_1)}\ ,
\label{4.1}
 \eeq
 where $g^h(x)$ is the gluon density in the incoming hadron $h$
 and
\beq
 \Delta x_1=\frac{\Delta E_G}{E_h}\ .
\label{4.2}
 \eeq
 
 However, with some maybe small probability $W_0$, the incoming
hadron
 may have no initial state inelastic interactions and produce
charmonium in
 the very first interaction without any energy loss. Thus the
nuclear
 suppression acquires two contributions \cite{kn1},
 \beqn
 Tr_A &=&
W_0
 \ +\ \frac{\sigma^{hN}_{in}}{A}\,
 \int d^2b\,
\int\limits_{-\infty}^\infty dz_2\,\rho_A(b,z)\,
 \nonumber\\ &\times&
\int\limits_{-\infty}^{z_2} dz_1\,
 \exp\left[-\sigma^{hN}_{in}
\int\limits_{-\infty}^{z_1} dz\,
 \rho_A(b,z)\right]\,
 R_G(x_1,\Delta x_1)\
.
\label{4.3}
 \eeqn
 Here the energy loss $\Delta E_G$ depends on the longitudinal
 distance
 $z_2-z_1$ between the first inelastic interaction and the
 production point
 of the charmonium. The probability of no initial state
 interaction reads,
 \beq
 W_0=\frac{1}{A\,\sigma_{in}^{hN}}\int
 d^2b\,\left[
 1-e^{-\sigma_{in}^{hN}\,T(b)}\right] =
 \frac{\sigma_{in}^{hA}}{A\,\sigma_{in}^{hN}}\ .
\label{4.4}
 \eeq
 The second term in (\ref{4.3}) is normalized to the probability
 $W_1=1-W_0$ of interaction in, the case of zero energy loss ($R_G=1$),
 \beq
 W_1=
 \frac{\sigma^{hN}_{in}}{A}
 \int
 d^2b\,\int\limits_{-\infty}^{\infty} dz_2\,
 \rho_A(b,z_2)\int\limits_{-\infty}^{z_2} dz_1\,
 \rho_A(b,z_1)\,{\rm
 exp}\left[-\sigma^{hN}_{in}
 \int\limits_{-\infty}^{z_1}
 dz\,\rho_A(b,z)\right]\ .
\label{4.5}
 \eeq
 
 In order to calculate nuclear suppression (\ref{4.3}) we need to
know
 the function $dE_G/dz$ which the rate of energy loss of a gluon
propagating through nuclear matter. Energy loss initiated by the
 inelastic
collision continues with a constant rate like it were in
 vacuum. One comes to
this conclusion both in the color-string model
 \cite{kn1,kn2} or treating
hadronization as gluon bremsstrahlung \cite{n,bh}.
 This so called vacuum energy
loss was supposed in \cite{kn1} to have a
 rate $-dE_G/dz=3\,GeV/fm$ leading to
a rather good description of data
 from the NA3 experiment \cite{na3}. A
substantially larger value
 $-dE_G/dz=5\,GeV/fm$ was advocated in \cite{hkz}
for the intermediate
 color-octet state in photoproduction of $J/\Psi$ off
nuclei.
 
 In fact, there is no controversy here, because in the case of
hadroproduction of charmonium it is more appropriate to rely upon the
 energy
loss of a quark, rather than a gluon, propagating through a
 medium. Indeed,
it was found in \cite{kst2} that light-cone gluons are
 located at small
transverse separation $\sim 0.3\,fm$ from the valence
 quarks.  This is
dictated by data for large mass diffraction which
 corresponds to diffractive
gluon radiation and was interpreted in
 \cite{kst2} as a result of a strong
nonperturbative interaction of the
 light-cone gluons. This observation is in
a good accord with other
 models for the nonperturbative QCD effects (see
discussion in
 \cite{k3p}). Thus, only a semi-hard interaction can resolve a
gluon in
 the constituent quark, while a soft interaction responsible for
inelastic interactions of the incident proton in the nucleus do not see
 the
gluon. The whole constituent quark, rather than the gluon, is
 hadronizing and
loosing energy propagating through the nucleus.
 Therefore, we should expect a
rate of energy loss similar to what is
 observed for Drell-Yan lepton pair
production off nuclei. The recent
 analysis \cite{eloss} of data \cite{e772}
for nuclear suppression in
 Drell-Yan reaction revealed for the first time a
nonzero energy loss for
 quarks, $-dE_q/dz=2.32 \pm 0.52 \pm
0.5\,GeV/fm$. This is compatible
 with the value $3\,GeV/fm$ used in
\cite{kn1,kn2}.

Thus, one can calculate the nuclear suppression of charmonium production
caused by energy loss in the same way as for Drell-Yan reaction 
\cite{eloss,eloss1},
 \beq 
R^{Eloss}_{A/N}(x_1) = W_0\ +\
\frac{\int\limits_{0}^{\infty} dL\,W_1(L)
\int\limits_{(x_q)_{min}}^{1}
d\,x_q\,F^h_q(x_q)\,\, d\,
\sigma^{qN}_{\chi}(\widetilde x_1^q)/ d\,
\widetilde x^q_1} {\int\limits_{x_1}^1
d\,x_q\,F^h_q(x_q)\,\, 
d\,\sigma^{qN}_{\chi}(x^q_1)/d\,x^q_1} \ .
\label{4.6}
 \eeq
 Here $F^h_q(x_q)$ is the quark distribution function in the
incident hadron, and $x_q$ and
$x^q_1=x_1/x_q$ are the fraction of the light-cone momentum of the incoming
hadron $h$ carried by the quark and the fraction of the quark momentum
carried by the $\bar cc$, respectively.  The lower integration limit
is given by $(x_q)_{min}=x_1+\Delta E/E_h$, and $\widetilde
x^q_1=x_1/(x_q-\Delta E/E_h)$.  The cross section 
$\sigma^{qN}_{\chi}(\widetilde x_1^q)$
of $\chi$ production
by a constituent quark, $q\,N \to \chi\,X$ is assumed to behave as
$\propto (1-x^q_1)^{2.5}$ in order to reproduce the observed $1-x_1)^5$
distribution of $J/\Psi$ production.

 At first glance the case of long coherence length is more
complicated
 since the $\bar cc$ fluctuation is produced long in advance and
propagates through the nucleus rather than the projectile gluon.
 However, the
soft interaction responsible for the first inelastic $h-N$
 collision (see
above) does not discriminate between the gluon and the
 color-octet $\bar
cc$. The same is true if instead of the light-cone
 representation one uses
the equivalent description of coherence via
 diffractive transition $G\to \bar
cc$ with no color exchange.
 Therefore, this inelastic interaction of the
projectile hadron and
 energy loss modify the ratio (\ref{4.2}) in the same
way as is described
 above for the limit of short coherence time. Thus,
Eq.~(\ref{4.3}) is
 valid in general case for any values of coherent and
formation lengths.

 Further, multiple interactions of a parton in a nuclear medium leads
 to a
broadening of the parton's transverse momentum and
 enhanced gluon
bremsstrahlung. The rate of associated induced energy
 loss rises linearly
with the length of the path in nuclear matter
 \cite{baier}, but is a rather
small correction to the vacuum energy
 loss.
 
 The induced energy loss
fluctuates and it becomes an important effect towards the
 kinematical limit $x_1\to 1$
where no gluon can be emitted. This of
 course suppresses the production rate
of charmonium, and more on a
 nuclear than on a proton target. Indeed, the
projectile gluon experiences broadening of the transverse momentum due to
multiple interactions (see in \cite{baier1} connection between $p_T$
broadening and induced energy loss). As a result gluon radiation becomes
more intensive. The
relation between induced energy loss and nuclear broadening of the mean  
transverse momentum squared of the quark was established in \cite{baier},  
 \beq 
\Delta E=\frac{3\,\alpha_s}{8}\, 
\Delta\la p_T^2\ra\,L\ ,
\label{4.8} 
 \eeq 
 where the broadening of $\la p_T^2\ra$ is proportional to
the length $L$ of the path in nuclear matter. The $p_T^2$ broadening for
$J/\Psi$ was measured in the E772 experiment 
\cite{psi-broad} to be rather small $\Delta\la p_T^2\ra\approx 0.5\,GeV^2$ even
for tungsten. Therefore the effective rate of induced energy loss in 
(\ref{4.8}) is
about $0.37\,GeV/fm$, nearly an order of magnitude smaller than the the value 
of vacuum energy loss we use. We assume that the induced energy loss is 
incorporated effectively in $dE/dz=-3\,GeV/fm$ we use in our calculations.

\section{Modification of the
\boldmath$x_F$-distribution by \boldmath$\chi\,\to\, J/\Psi\,\gamma$
 decay}\label{decay}

No data is still available for $x_F$-distribution of produced $\chi$, but
only for $J/\Psi$. About $40\%$ of them originate from the $\chi\to
J/\Psi\,\gamma$ decay and have momenta smaller than that of $chi$.
 Therefore
we should correct the $\chi$ $_F$ distribution calculated in
 previous section
to compare with data for indirect $J/\Psi$.
 It would be incorrect, however,
to assume isotropic decay of $\chi$
 since it is produced polarized.
 
The structure of the amplitude of decay of the tensor meson ($\chi_2$) to two
vector mesons with masses $m_1$ and $m_2$ is fixed by the gauge
 invariance,
\beq
 A(T\to V_1+V_2) \propto
 h^{\mu\nu}\left\{\left[\tilde e^\mu_1 -
\frac{(\tilde e_1p)p^\mu}{p^2}\right]
 \left[\tilde e^\mu_2 -
 \frac{(\tilde
e_2p)p^\mu}{p^2}\right]+
 \Bigl(\mu\to\nu\Bigr)\right\}\ .
\label{5.1}
 \eeq
 Here
 \beqn
 \tilde e^\mu_1 &=&
 e^\mu_1 - \frac{(e_1k_2)k^\mu_1}
 {k_1k_2}\ ;\nonumber\\
 \tilde e^\mu_2 &=&
 e^\mu_2 -
 \frac{(e_2k_1)k^\mu_2}
 {k_1k_2}\ ,
\label{5.2}
 \eeqn
 where $e_{1,2}$ are the polarization vectors of $V_{1,2}$,
 \beq
e_1k_1=e_2k_2=0\ ;
\label{5.3}
 \eeq
 $p$ is the 4-momentum of the $\chi_2$; $k_{1,2}$ are the 4-momenta of
$V_1$ and $V_2$ respectively;
 $h^{\mu\nu}=h^{\nu\mu}$ is the $\chi_2$
polarization tensor satisfying
 the conditions,
 \beqn
 g_{\mu\nu}h^{\mu\nu}
&=& 0\ ;
 \nonumber\\
 p_\mu h^{\mu\nu} &=& 0\ .
\label{5.4}
 \eeqn
 In the rest frame of the $\chi_2$ the
 components of the polarization
tensor $h^{00}=h^{0i}=h^{i0}=0$.
 Other components for the state with the
projection $m=2$ are,
 \beq
 h^{ik}(m=2) = \epsilon_+^i\epsilon_+^k\ ,
\label{5.5}
 \eeq
 where
 \beq
 \vec\epsilon_+ =
 -\,\frac{\vec e_x+\vec
e_y}{\sqrt{2}}\ .
\label{5.6}
 \eeq
 
 The angular distribution of the decay products relative to the
momentum of $\chi_2$ has the form,
 \beq
 \frac{d\,N}{d\,\cos\theta}
\propto
 (1 + \beta_1\,\cos^2\theta)
 (1 + \beta_2\,\cos^2\theta)\ ,
\label{5.7}
 \eeq
 where $\beta_{1,}$ are the velocities of the produced $V_{1,2}$.
 In
 the case of $V_1=J/\Psi$, $V_2=\gamma$ we have $\beta_1\approx
 1/9$,
 $\beta_2=1$.
 
\section{Comparison with available data and predictions for higher 
energies}\label{results}

\begin{figure}[tbh]
\includegraphics{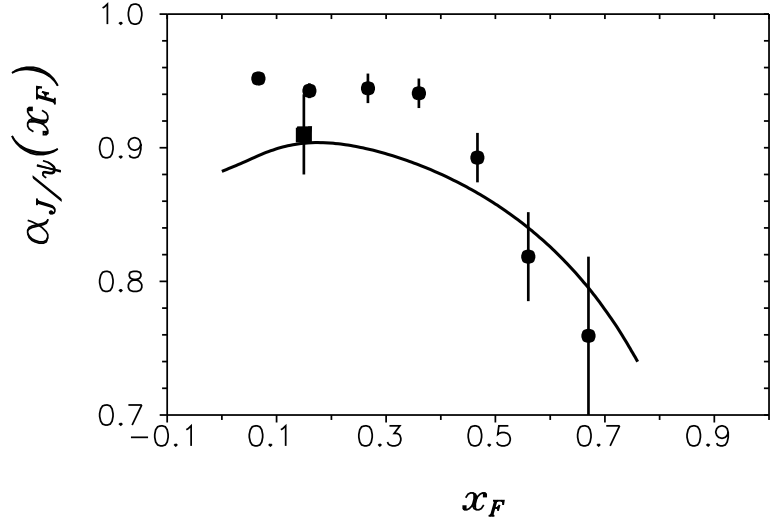}
\begin{center}
\vspace{7cm}
\parbox{13cm}
{\caption[Delta]
{\sl $x_F$ dependence of nuclear suppression of 
$J/\Psi$ production at $200\,GeV$ in terms of
the exponent $\alpha$ characterizing the $A^\alpha$ dependence of
$J/\Psi$ production. Energy loss Eq.~(\ref{4.6}), absorption, 
quark shadowing and gluon enhancement are included
in the calculations. The round data points are from 
\cite{na3}, the square point is from \cite{na38-alpha} .}
\label{na3}}
\end{center}
 \end{figure}

First of all, we checked with data at the energies of the SPS where the 
coherence length is rather short and the 
main effects are absorption and energy loss.
With (\ref{4.6}) we calculated the $x_F$ dependence of the exponent 
$\alpha(x_F)$ describing the $A$-dependence of the 
$J/\Psi$ production rate, $\propto A^{\alpha(x_F)}$  
at $200\,GeV$. We also corrected it for gluon enhancement
and $\chi\to J/\Psi\,\gamma$ decay. 
The results are compared with data \cite{na3} in Fig.~\ref{na3}.
The observed steep fall of $\alpha$ is well reproduced, while at small $x_F$
our calculations seem to underestimate the data from the NA3 experiment. 
In fact, our calculations for energy loss effect are not 
trustable at small $x_F$.
Indeed, the suppression caused by the shift of energy depends on how steep
the $x^q_1$ dependence of the cross section $qN\to \chi X$ is, the steeper it is, 
the stronger is the suppression. The chosen parametrization
$d\sigma^{qN}_{\chi}/d\,x^q_1 \propto (1-x^q_1)^{2.5}$
is valid only at large $x^q_1\to 1$. This DY cross section in $NN$ collision
must be flat around $x_F=0$, what makes the
suppression by energy loss weaker at small $x_F$.

At higher energy, $800\,GeV$ the dynamics of $J/\Psi$ suppression 
is more complicated and includes more effects.
 Now we can apply more corrections to the dotted curve in
Fig.~\ref{xf-dep} which involves only quark and gluon shadowing. Namely,
inclusion of the energy loss effect and decay $\chi\to J/\Psi\,\gamma$
leads to a stronger suppression depicted by dashed curve. Eventually we
correct this curve for gluon enhancement at $x_2\sim 0.1$ (small $x_F$)
and arrive at the final result shown by thick solid curve.

Since our calculation contains no free parameters we think that the 
results agree with the data amazingly well. Some difference in the shape 
of the maximum observed and calculated at small $x_F$ may be a result of 
the used parameterization \cite{ekr} for gluon antishadowing. We think that it
gives only the scale of the effect, but neither the ad hoc shape, 
nor the magnitude should be taken literally.
Besides, our calculations are relevant only for those 
$J/\Psi$s which originate from $\chi$ decays which feed only about $40\%$ 
of the observed ones.

\begin{figure}[tbh]
\includegraphics{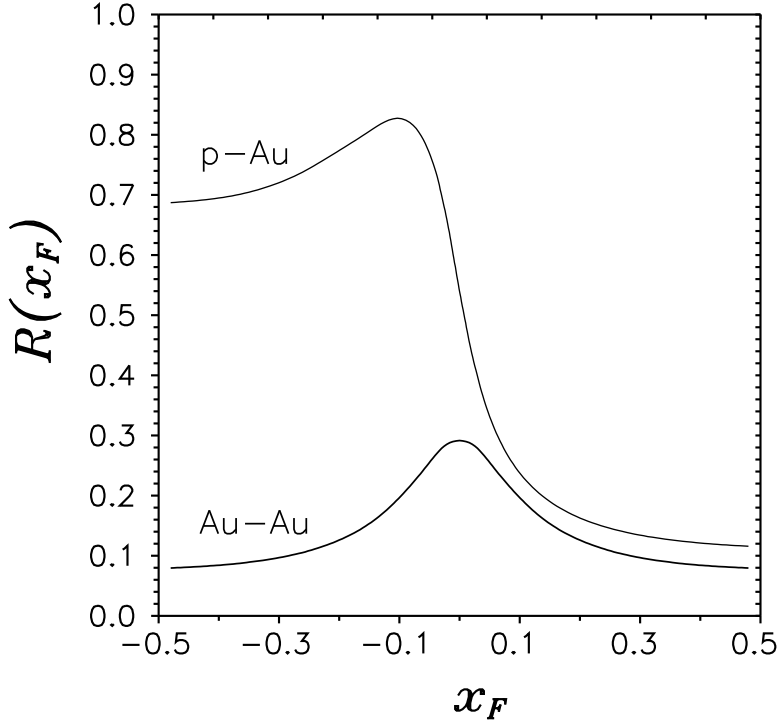}
\begin{center}
\vspace{10cm}
\parbox{13cm}
{\caption[Delta]
{\sl Nuclear suppression of $J/\Psi$ production in proton-gold 
collisions at $\sqrt{s}=200\,GeV$ as function of $x_F$ (the upper curve)
and in gold-gold collisions (bottom curve). Effects of quark and gluon 
shadowing and gluon antishadowing are included.}
\label{rhic}}
\end{center}
 \end{figure}

At higher energies of the RHIC and LHC the effect of energy loss is 
completely gone and nuclear suppression must expose the $x_2$ scaling.
Much smaller $x_2$ can be reached at higher energies. 
Our predictions for proton-gold to proton-proton ratio is depicted in 
Fig.~\ref{rhic}. 
One can see that at $x_F > 0.1$ shadowing suppresses charmonium 
production by nearly an order of magnitude.

We can also estimate the effect of nuclear suppression in heavy ion 
collisions assuming factorization,
 \beq
R_{AB}(x_F) = R_{pB}(x_F)\,R_{pA}(-x_F)\ .
\label{AB}
 \eeq
Our predictions for gold-gold collisions at $\sqrt{s}=200\,GeV$ are 
shown by the bottom curve in Fig.~\ref{rhic}. Since factorization is 
violated this prediction should be verified.

\section{Conclusions and outlook}\label{conclusions}

 The long standing challenge of explanation of the steep $x_F$ 
dependence of charmonium suppression observed in
$pA$ collisions becomes an appealing problem with advent of RHIC
and LHC colliders. A good understanding of the underlying mechanisms is
especially important since  charmonium suppression is one of
the major probes for the quark-gluon plasma formation in heavy ion collisions. 
The conventional mechanisms of charmonium suppression used as a
baseline for new physics in experiments at SPS CERN must be essentially
revised. This demands development of new theoretical tools, and first of all
an explanation for already collected data for charmonium suppression.
 
Coherent effects or shadowing become the dominant effect which governs 
charmonium suppression at RHIC. So far no theoretical tool has been
available to deal with this phenomenon. The main result of present paper is 
the development of the light-cone dipole approach describing 
coherence effects at any energy. It can be structured as follows.

\begin{itemize}

\item
Final state absorption of the produced colorless $\bar cc$ in nuclear matter is 
accomplished by initial state shadowing at high energies when the coherence time
Eq.~(\ref{2}) is sufficiently long. A light-cone Green function formalism
is developed which describes propagation of fluctuating color octet (singlet)
$\bar cc$ wave packets
through nuclear matter. This approach includes both the effects of
formation and coherence. It interpolates between the low energy limit, $t_f\ll R_A$,
where only final state absorption causes charmonium suppression, and the high 
energy limit, $t_c\gg R_A$ where the color octet $\bar cc$ fluctuation is 
developed long in advance and propagates through the whole nuclear thickness
and is converted to a color singlet. This is an analog to shadowing of $c$ quarks
in DIS. Its existence is a deviation from QCD factorization which is quite strong,
an about $50\%$ effect on heavy nuclei.

\item
At high energies fluctuations are overwhelmed by gluons which are the source
of rising energy dependence of all cross sections. Such fluctuations are
rather heavy and their lifetime is much shorter than that of a $\bar cc$ pair.
Only at very high energies these gluonic fluctuations live sufficiently
long to participate in multiple interactions which lead to shadowing.
This gluon shadowing turns out to be very strong and becomes the main source of 
charmonium suppression at RHIC. Again QCD factorization is strongly violated.
Namely, gluon shadowing for colorless $\bar cc$ production turns out to be
much stronger than in DIS of open charm production.

\item
Both mechanisms, shadowing of $c$ quarks and gluons depend only on $x_2$.
Although available data at medium high energies strikingly contradict 
$x_2$ scaling, it comes from energy loss effects. Indeed, modification
of the $x_1$ distribution by the energy loss depends both on $x_1$ and
incident energy. Accidentally, data demonstrate an approximate $x_F$ scaling
in the energy interval $200-800\,GeV$ which our calculations confirm.
It is a result of interplay between energy loss and shadowing: the former
effects vanishing with energy, while the latter is rising.
Exact $x_2$ scaling is predicted at high energies (well beyond $1\,TeV$)
of RHIC and LHC. We predict a very steep variation of charmonium suppression
around $x_F=0$ at RHIC. In fact, it can also be obtained directly from the 
E866 data neglecting energy loss corrections at $800\,GeV$. For gold-gold
collisions we predict a peculiar narrow peak at $x_F=0$.

\end{itemize}

 As a further development we are going to extend this approach
to the case of directly produced $J/\Psi$ and $\Psi^\prime$.
We do not expect the results to be very different what what we found in this paper.
Indeed, the kinematics of decays $\chi \to J/\Psi\,\gamma$ and
$|\bar cc\ra_8 \to J/\Psi\,G$ is similar up to the difference in the
invariant masses of the $\bar cc$ pairs. However, we demonstrate in Sect.~\ref{decay}
that the decay has practically no effect, and the $x_F$ distributions 
of $\chi$ and $J/\Psi$ are nearly identical.

Another source of direct production is hidden in
the nuclear effects for $\chi$ production. Namely,
the $\bar cc$ pair traveling through nuclear matter and experiencing
multiple interactions changes its spin-orbital structure, therefore it also
feeds the $\Psi$ channel. We eliminated this channel in Sect.~\ref{quarks}
projecting the $\bar cc$ on the state with the quantum numbers of $\chi$.
If to project the $\bar cc$ on the $\Psi$ one gets a cross section comparable with
data. Therefore, this channel of direct $\Psi$ production is important and should
be studied. 

It is still a challenge how to apply the developed approach to
nucleus-nucleus collisions. The relation Eq.~(\ref{AB}) between nuclear effects
in $pA$ and $AA$ collisions is just an approximation based on QCD factorization 
which as we saw is badly violated. We also hope to make a progress in this
direction.

\medskip

{\bf Note added:} 
Soon after this paper has been submitted to the Los Alamos bulletin board, another
paper was put on the server \cite{nantes1} which focuses at the same problem
of charmonium production at the energies of RHIC. Nuclear suppression at RHIC is
predicted only at very small $x_F < 0.084$ assuming $x_2$ scaling and employing the
E866 data \cite{e866}. At larger $x_F$ nuclear effects are calculated within
model \cite{nantes2} fitted to the E866 data. The model 
has enough free parameters to be successful with available data, but it 
completely misses the coherence effects, in particular  
gluon shadowing, which are the main source of $J/\Psi$ suppression 
at high energies. Therefore, extrapolation to higher energies beyond the
$x_2$ range of the E866 data should not be trusted.

On the other hand, we do not rely on either 
$x_2$ scaling (which is still not at work at $800\,GeV$,
see Fig.~\ref{xf-dep}), or the E866 data,
but directly calculate nuclear effects at the energy of RHIC.
We compare our parameter free calculations with the E866 data only to 
demonstrate agreement.

\bigskip 
\noindent {\bf Acknowledgment}: this work has been partially supported by the
grant from the Gesellschaft f\"ur Schwerionenforschung Darmstadt (GSI), grant
No.~GSI-OR-SCH, and by the Federal Ministry BMBF grant No.~06 HD 954.
A part of this work was done when A.V.T. was employed by Heidelberg
University and his work has been supported by the grant from the Gesellschaft
f\"ur Schwerionenforschung Darmstadt (GSI), grant no. HD~H\"UFT.

 \def\appendix{\par
 \setcounter{section}{0}
\setcounter{subsection}{0}
 \def\thesection{Appendix \Alph{section}}
\def\thesubsection{\Alph{section}.\arabic{subsection}}
\def\theequation{\Alph{section}.\arabic{equation}}
\setcounter{equation}{0}}

 \appendix
 
\section{LC wave functions of the \boldmath$\chi$ states}
\setcounter{equation}{0}
 
 We are going to establish relation between
 the
wave functions of the $^3P_J$ states ($\chi_J$
 mesons) in the LC and rest
frames.
 First, the LC wave function in coordinate space,
 \beq
\Phi_{\chi}^{\mu\bar\mu}(r_T,\alpha)=\frac{1}{2\pi}\int
 d^2p_T\,e^{-i\vec
p_T\cdot\vec r_T}\,
 \Phi_{\chi}^{\mu\bar\mu}(p_T,\alpha)\ ,
\label{a.1}
\eeq
 is normalized as,
 \beq
 \sum\limits_{\mu\bar\mu}
 \int\limits_0^1d\alpha\int d^2r_T\,
 \Bigl|\Phi_{\chi}^{\mu\bar\mu}(r_T,\alpha)\Bigr|^2 =
 \sum\limits_{\mu\bar\mu}
 \int\limits_0^1d\alpha\int d^2p_T\,
 \Bigl|\Phi_{\chi}^{\mu\bar\mu}(p_T,\alpha)\Bigr|^2 =
 1\ .
\label{a.2}
\eeq
 The problem of Lorentz boosting of a quarkonium wave function
 from the
rest frame, where it can be found as a solution of
 the Schr\"odinger
equation, to the LC frame still has no well grounded
 solution. A
nonrelativistic two-body system $\bar cc$ in its rest
 frame corresponds to
the full series of Fock states, which include
 gluons and the sea, in the LC
frame.
 To construct the wave function for the lowest Fock component $\bar
cc$
 we use the popular recipe suggested in \cite{teren} (see also in
\cite{hp,hikt}),
 \beq
 \Phi_{\chi}^{\mu\bar\mu}(p_T,\alpha) =
\left(\frac{\partial\,p_L(p_T,\alpha)}
{\partial\,\alpha}\right)^{1\over2}\,
 \Psi_{\chi}^{\mu\bar\mu}(\vec p)
\Bigr|_{\vec p=\vec p_T+p_L\vec n}\ ,
\label{a.3}
\eeq
 where
 \beqn
 &&p_L = \left(\alpha-1/2\right)\,
 M(p_T,\alpha)\ ,\nonumber\\
 &&M^2(p_T,\alpha) = \frac{m_c^2+p_T^2}
 {\alpha(1-\alpha)}\ .\nonumber
\label{a.4}
\eeqn
 The rest frame wave function $\Psi_{\chi}^{\mu\bar\mu}(\vec p)$
 for a state with the total angular momentum $J$ and projection $J_z$
 on the
 direction $\vec n$ is a combination of different spin and
 orbital momenta,
 \beq
 \Psi_{\chi}^{\mu\bar\mu}(\vec p) =
 \sum\limits_{m_1+m_2=J_z}
 \la 1,1;J,J_z|1,1;m_1,m_2\ra\,
 S_{1,m_1}^{\mu\bar\mu}\,
 \psi_{1,m_2}(\vec
 p)\,
\label{a.5}
\eeq
 where the orbital part $\psi_{1,m_2}^{\mu\bar\mu}(\vec p)$
 is a
 Fourier transform of the spatial part of the wave function,
 \beq
 \psi_{1,m_2}(\vec p)=
 \frac{1}{(2\pi)^{3/2}}\,
 \int d^3r\,e^{i\vec
 p\cdot\vec r}\,
 \psi_{1,m_2}(\vec r)\ .
\label{a.6}
\eeq
 The latter, as usual, can be represented as a product of the
 angular
 and radial $r$-dependent parts,
 \beq
 \psi_{1,m_2}(\vec r)
 =
 Y_{1,m_2}\left(\vec r/r\right)\,
 R_1(r)\ ,
\label{a.7}
\eeq
 It is normalized as,
 \beq
 \int d^3r\,\Bigl|\psi_{1,m_2}(\vec
 r)\Bigr|^2 =
 \int dr\,r^2\,R^2_1(r) = 1\ .
\label{a.8}
\eeq
 
 The orbital part has the form,
 \beq
 Y_{1,m_2}\left(\vec
r/r\right) =
 \sqrt{\frac{3}{4\pi}}\,
 \frac{\vec e_{m_2}\cdot\vec r}{r}\ ,
\label{a.9}
\eeq
 where the polarization vectors are defined as,
 \beq
 \vec e_m
 = \left\{
\begin{array}{ccc}
\vec e_+=-(\vec e_x+i\vec e_y)/\sqrt{2} &
 \ \ \ \ & m=1\\
 \vec e_0=\vec
e_z=\vec n &
 \ \ \ \ & m=0\\
 \vec e_+=(\vec e_x-i\vec e_y)/\sqrt{2} &
 \ \
\ \ & m=-1
\end{array}\right.\ .
\label{a.10}
\eeq
 They satisfy the relations,
 \beqn
 \vec e_m\cdot\vec e_k &=&
 \delta_{mk}\ ,\nonumber\\
 \vec e_+\cdot\vec e_- &=& -1\ ,\nonumber\\
 \vec
 e_+\cdot\vec e_+ &=&
 \vec e_-\cdot\vec e_- =
 \vec e_+\cdot\vec e_0 =
 \vec e_-\cdot\vec e_0 = 0\ .
\label{a.11}
\eeqn
 
 The spin part reads,
 \beq
 S_{1,m_1}^{\mu\bar\mu} =
\frac{1}{\sqrt{2}}\,
 \Bigl.\xi_c^\mu\Bigr.^\dagger\,
 \vec\sigma\cdot\vec
e_{m_1}\,
 \tilde\xi_{\bar c}^{\bar\mu}\ ,
\label{a.12}
\eeq
 where
 \beq
 \tilde\xi_{\bar c}^{\bar\mu}=
 i\,\sigma_y\,\xi_{\bar c}^{\bar\mu}\Bigr.^*\ ,
\label{a.13}
\eeq
 and spinors $\xi$ in the charmonium rest frame are related to those,
 $\eta$, in the LC frame by the Melosh spin rotation transformation,
 \beqn
 \xi_c^\mu &=& \hat R(\vec p_T,\alpha)\,\eta_c^{\mu}\ ,
 \nonumber\\
 \xi_{\bar c}^{\bar\mu} &=&
 \hat R(-\vec p_T,1-\alpha)\,\eta_{\bar
 c}^{\bar\mu}\ .
\label{a.14}
\eeqn
 The rotations matrix reads,
 \beq
 \hat R(\vec p_T,\alpha)
 =
 \frac{m_c+\alpha\,M -
 i\vec\sigma\cdot(\vec n+\vec p_T)}
 {\sqrt{(m_c+\alpha\,M)^2+p_T^2}}
 \ .
\label{a.15}
\eeq
 
 Then spin part of the LC wave function takes the form,
 \beq
S_{1,m_1}^{\mu\bar\mu} =
 \frac{1}{\sqrt{2}}\,\eta_c^{\mu}\Bigr.^\dagger\,
\hat A_{m_1}(\vec p_T,\alpha)\,
 \tilde\eta_{\bar c}^{\bar\mu}\ ,
\label{a.16}
 \eeq
 where
 \beq
 \hat A_{m_1}(\vec p_T,\alpha) =
 \hat R^\dagger(\vec
p_T,\alpha)\,
 (\vec\sigma\cdot\vec e_{m_1})\,
 \sigma_y\,\hat R^*(\vec
p_T,\alpha)\,
 \sigma_y^{-1} \equiv \hat B_{m_1} + \hat C_{m_1}\ ,
\label{a.17}
 \eeq
 \beq
 \hat B_{m_1} = \frac{1}{m_T}\,
 \Bigl[m_c\,\vec\sigma\cdot\vec
 e_{m_1} +
 i(\vec e_{m_1}\times\vec n)\cdot\vec p_T\Bigr]\ ,
\label{a.18}
 \eeq
 \beq
 \hat C_{m_1} =
 2\,\frac{(\vec e_{m_1}\cdot\vec p_T)
 (\vec\sigma\cdot\vec p_T-p_L\,\vec\sigma\cdot\vec n)
 + (\vec
 e_{m_1}\cdot\vec n)
 (p_L\,\vec\sigma\cdot\vec p_T +
 p_T^2\,\vec\sigma\cdot\vec n)}
 {m_T(2\,m_c + M)}\ ,
\label{a.19}
 \eeq
 and $m_T=\sqrt{m_c^2+p_T^2}$.

\section{The \boldmath$\alpha$ distribution of the charmonium wave function}
\setcounter{equation}{0}
 
 The $\alpha$-dependence of the charmonium LC wave
function
 picks at $\alpha=1/2$ and has a width related to the
 mean
longitudinal velocity $\la v_L^2\ra$ of the quarks,
 \beq
\left\la\left(\alpha-{1\over2}\right)^2\right\ra =
 \frac{\la
p_L^2\ra}{4\,m_c^2} =
 {1\over4}\,\la v_L^2\ra\ ,
\label{bb.1}
 \eeq
 which is a rather small for nonrelativistic systems.
 
 The relative
$\bar cc$ velocity can be estimated applying the Virial
 theorem to the
realistic power-law potential
 $V(r)=-V_0+V_1(r/r_0)^\lambda$ which describes
well the properties of
 charmonia \cite{power}. Then, the mean kinetic energy
$\la E_k\ra=m_c\la
 v^2\ra$ is related to the power $\lambda$,
 \beq
 \la
E_k\ra = \frac{\lambda}{2+\lambda}\,
 \Bigl(M_\Psi-2m_c+V_0\Bigr)\ ,
\label{bb.2}
 \eeq
 where $M_\Psi$ is the mass of the charmonium. Using $\lambda=0.1$ and
 $V_0=8\,GeV$ \cite{power} we get $\la v^2\ra = 0.21$. Since $V_0$ is so
 large the mean velocity turns out to be nearly the same for all charmonia
 ($J/\Psi,\ \Psi',\ \chi$. Numerical calculations \cite{hikt} confirm this,
 and also show that the results is the same for all realistic potentials.
 This is not surprising since all those potentials look pretty similar (see
 in \cite{hikt}).
 
 Note that the $\chi$ states with projections $J_z=\pm 1$
 are not produced
 since the gluons are transversely polarized. The $\chi_2$
 with $J_z=0$ is
 not produced at all in the nonrelativistic limit $v\to 0$
 \cite{bhtv} and
 is suppressed by the tiny factor $v^4$. Therefore, only the
 $\chi_2$ with
 projection $J_z=\pm 2$ is of practical interest.
 
 For the
 S-wave state the mean longitudinal speed squared in
 Eq.~(\ref{bb.1}) would
 be $\la v_L^2\ra=\la v^2\ra/3$. However, for the
 P-wave state with
 $J_z=\pm2$ we are interested in, the corresponding
 factor is even smaller,
 $\la v_L^2\ra=\la v^2\ra/5$, leading to a very
 small value of
 $\la(\alpha-1/2)^2\ra = \la v^2\ra/20 = 0.01$.
 
\section{Landau-Yang theorem: manifestation of relativistic effects}
\setcounter{equation}{0}
 
 Obviously, the spin rotation formalism is rather
complicated, and one
 might hope that for nonrelativistic systems like
charmonium the
 corrections are small since $\hat R(\vec p_T,\alpha)\to 1$
when $v\to
 0$. In many cases it is indeed true, for instance the cross
section of
 charmonium photoproduction increases by $30\%$ (not a negligible
correction either) when the spin-rotation effects are included
\cite{hikt}. In some cases, however, the spin rotation is very
important. Photoproduction of the $2S$ state $\Psi^\prime$, for
 instance, is
enhanced by factor $2-3$ \cite{hikt}. In the case under
 discussion of
hadroproduction of $\chi$ states the spin rotations also
 plays a principal
role. Its suppression leads to a rather intensive
 production of $^3P_1$ even
by two massless gluons forbidden by the
 Landau-Yang theorem.  We demonstrate
here that inclusion of the
 spin-rotation resolves this controversy.
 
Since production of any state with projection $J_z=\pm1$ by two
 on-mass-shell
gluons is forbidden we should consider only the $\chi_1$
 state with
$J_z=0$. Its wave function in the rest frame is a linear
 combination of the
spin and orbital parts with known Clebsch-Gordan
 coefficients,
 \beq
\Psi^{\mu\bar\mu}_{\chi}(\vec p)=
 \frac{1}{\sqrt{2}}\,
\Bigl[S^{\mu\bar\mu}_{+1}\,\psi_{1,-1}(\vec p) -
S^{\mu\bar\mu}_{-1}\,\psi_{1,+1}(\vec p)\Bigr]\ .
\label{b.1}
 \eeq
 Substituting here Eqs.~(\ref{a.16})-(\ref{a.19}) we get,
 \beq
\Psi^{\mu\bar\mu}_{\chi}(\vec p)=
 {1\over2}\,\eta^{\mu}_c\Bigr.^+\,
\Bigl[(\hat B_{+1}+\hat C_{+1})\,\psi_{1,-1}(\vec p)-
 (\hat B_{-1}+\hat
C_{-1})\,\psi_{1,+1}(\vec p)\Bigr]\,
 \tilde\eta^{\bar\mu}_{\bar c}\ .
\label{b.2}
 \eeq
 This expression can be simplified using relations,
 \beq
\psi_{1,m}(\vec p)=
 -i\,\vec e_m\cdot\vec p\,R_1(\vec p)\ ,
\label{b.3}
 \eeq
 where
 \beq
 R_1(\vec p) = \int\limits_0^\infty dr\,
 \left({r\over
p}\right)^{{3\over2}}\,
 J_{3/2}(pr)\,R_1(r)\ ,
\label{b.4}
 \eeq
 and also,
 \beq
 \hat C_{\pm1} = \frac{2\,(\vec e_{\pm1}\cdot\vec
p)\,
 (\vec\sigma\cdot\vec p_T -
 p_L\,\vec\sigma\cdot\vec n)}
{m_T\,(2m_c+M)}\ .
\label{b.5}
 \eeq
 The terms proportional to $\hat C$ cancel and the wave function
 of
the state with $J=1$, $J_z=0$ gets the simple form,
 \beqn
\Psi^{\mu\bar\mu}_{\chi_1}(\vec p) &=&
 -\frac{i\,R_1(\vec
p)}{2\,m_T}\,\eta_c^\mu\Bigr.^\dagger\,
 \left\{\Bigl[m_c\,\vec\sigma\cdot\vec
e_+ +
 i(\vec e_+\times\vec n)\cdot\vec p_T\Bigr]\,
 \vec e_-\cdot\vec p_T
\right.\nonumber\\
 &-& \left.\Bigl[m_c\,\vec\sigma\cdot\vec e_- +
 i(\vec
e_-\times\vec n)\cdot\vec p_T\Bigr]\,
 \vec e_+\cdot\vec p_T
 \right\}\,
\tilde\eta_{\bar c}^{\bar\mu}\ .
\label{b.6}
 \eeqn
 
 Using this expression we eventually arrive at the LC wave
function which has the form,
 \beqn
 \Phi^{\mu\bar\mu}_{\chi_1}(\vec
r_T,\alpha) &=&
 {1\over2}\,\eta_c^\mu\Bigr.^\dagger\,
\left\{\Bigl[m_c\,\vec\sigma\cdot\vec e_+ +
 i(\vec e_+\times\vec
n)\cdot\vec\nabla\Bigr]\,
 \vec e_-\cdot\vec r_T
 \right.\nonumber\\
 &-&
\left.\Bigl[m_c\,\vec\sigma\cdot\vec e_- +
 i(\vec e_-\times\vec
n)\cdot\vec\nabla\Bigr]\,
 \vec e_+\cdot\vec r_T
 \right\}\,H(r_T)\,
\tilde\eta_{\bar c}^{\bar\mu}\ ,
\label{b.7}
 \eeqn
 where
 \beq
 H(r_T) = \frac{1}{m_T\,r_T}\,
\int\limits_{0}^{\infty} dp_T\,
 p_T^2\,J_1(p_Tr_T)\,\left(\frac{\partial
p_L}
 {\partial\alpha}\right)^{1\over2}\,R_1(\vec p)\ .
\label{b.8}
 \eeq
 
 Eq.~(\ref{b.7}) can be further simplified taking into account
relations,
 \beqn
 &&
 (\vec\sigma\cdot\vec e_+)\,\vec e_- -
(\vec\sigma\cdot\vec e_-)\,\vec e_+ =
 \vec\sigma\times(\vec e_-\times\vec
e_+)\ ,
 \nonumber\\
 &&
 \vec e_+\times\vec e_-=i\vec n\ ,\ \ \ \ \
 \vec
e_+\cdot\vec e_-= -1
 \nonumber\\
 &&
 \vec e_+\times\vec n=i\vec e_+\ ,\ \
\ \ \
 \vec e_-\times\vec n=i\vec e_-\ ,
 \nonumber\\
 &&
 (\vec
e_+\cdot\vec r_T)(\vec e_-\cdot\vec r_T)=
 -{r_T^2\over2}\ .
\label{b.9}
 \eeqn
 The final result reads,
 \beq
 \Phi^{\mu\bar\mu}_{\chi_1}(\vec
r_T,\alpha) =
 -{i\over2}\,\eta_c^\mu\Bigr.^\dagger\,
\left\{(\vec\sigma\times\vec n)\cdot\vec r_T\,H(r_T) +
 {1\over
r_T}\,\frac{d}{d\,r_T}\,\Bigl[r_T^2\,H(r_T)\Bigr]\,
\right\}\,\tilde\eta_{\bar c}^{\bar\mu}\ .
\label{b.10}
 \eeq
 
 Now let us consider the function,
 \beq
 A =
\sum\limits_{\mu,\bar\mu}\,
 \int\limits_0^1 d\alpha \int d^2r_T\,
\Phi^{\mu\bar\mu}_{\chi_1}(\vec r_T,\alpha)\Bigr.^*\,
 (\vec e_T\cdot\vec
r_T)\,
 \Phi^{\mu\bar\mu}_{G}(\vec r_T,\alpha)\ ,
\label{b.11}
 \eeq
 which coincides with the amplitude of $\chi_1$ production in
gluon-nucleon collision up to the
 factor $\sqrt{C(s)/8}$, if the
approximation of dipole cross section
 $\sigma_{\bar qq}(r_T,s) = C(s)\,r_T^2$
is applied.
 This amplitude can be represented as,
 \beq
 A=A_1+A_2\ ,
\label{b.12}
 \eeq
 where
 \beq
 A_1 = i\,m_c^2\,\int\limits_0^1 d\alpha\int d^2r_T\,
(\vec e\times\vec n)\cdot\vec r_T\,
 H(r_T)\,(\vec e\cdot\vec
r_T)\,K_0(\epsilon r_T)\ ;
\label{b.13}
 \eeq
 \beqn
 A_2 &=& i\,\int\limits_0^1 d\alpha\int d^2r_T\,
 {1\over
 r_T}\,\frac{d}{d\,r_T}\,
 \Bigl[r_T^2\,H(r_T)\Bigr]\,
 (\vec e_T\cdot\vec
 r_T)\,
 (\vec e\times\vec n)\cdot\vec\nabla\,
 K_0(\epsilon r_T)\nonumber\\
 &=&
 -i\,\epsilon\,\int\limits_0^1 d\alpha\int d^2r_T\,
 {1\over
 r_T^2}\,\frac{d}{d\,r_T}\,
 \Bigl[r_T^2\,H(r_T)\Bigr]\,
 (\vec e_T\cdot\vec
 r_T)\,
 (\vec e\times\vec n)\cdot\vec r_T\,
 K_1(\epsilon r_T)\ ,
\label{b.13a}
 \eeqn
 using the general property of the spinors $\eta$,
 \beq
\sum\limits_{\mu,\bar\mu}
 \Bigl(\eta_{c}^{\mu}\Bigr.^\dagger\,
 \hat
a\,\tilde\eta_{\bar c}^{\bar\mu}\Bigr)^*\,
\Bigl(\eta_{c}^{\mu}\Bigr.^\dagger\,
 \hat b\,\tilde\eta_{\bar
c}^{\bar\mu}\Bigr) =
 {\rm Tr}\Bigl(\hat a^\dagger\hat b\Bigr)\ ,
\label{b.14}
 \eeq
 where $\hat a$ and $\hat b$ are any two-dimensional matrices.
 
Integration over azimuthal angle in Eqs.~(\ref{b.12})-(\ref{b.13a})
 can be be
performed using relation,
 \beq
 \int\limits_{0}^{2\pi} d\phi\,
r_{T}\Bigr._{i}\,r_{T}\Bigr._{k} =
 \pi\,\delta_{ik}\,r_T^2\ ,
\label{b.15}
 \eeq
 and we arrive at the following common form for $A_1$ and $A_2$,
\beq
 A_{1,2} = i\,\pi\,(\vec e\times\vec e_T)\cdot\vec n\,
 \int\limits_0^1
d\alpha\,I_{1,2}(\alpha)\ ,
\label{b.16}
 \eeq
 where
 \beqn
 I_1(\alpha) &=& m_c^2\,\int\limits_0^\infty
dr_T\,r_T^3\,K_0(\epsilon r_T)\,H(r_T)\ ,
\label{b.17}\\
I_2(\alpha) &=& -\epsilon \int\limits_0^\infty
 dr_T\,r_T\,K_1(\epsilon
r_T)\,
 \frac{d}{d\,r_T}\Bigl[r_T^2\,H(r_T)\Bigr]\ .
\label{b.18}
 \eeqn
 Integration in the last equation can be performed by parts
 using the
relation,
 \beq
 \frac{d}{d\,r_T}\,r_T\,K_1(\epsilon r_T) =
 -
\epsilon\,r_T\,K_0(\epsilon r_T)\ .
\label{b.19}
 \eeq
 The result reads,
 \beq
 I_1(\alpha) =
-\epsilon^2\,\int\limits_0^\infty
 dr_T\,r_T^3\,K_0(\epsilon r_T)\,H(r_T)\ ,
\label{b.20}
 \eeq
 
 Thus, we conclude that in the limit of on-mass-shell gluon,
($\epsilon\to m_c$), $I_2\to-I_1$, and the amplitude (\ref{b.12}) of
 $\chi_1$
production vanishes in accordance with the Landau-Yang
 theorem. This is not
an obvious result, indeed, the procedure
 of of Lorentz boost to the infinite
momentum frame for the
 quarkonium wave function is ill defined, as was
mentioned above.
 It is important that it survives such a rigorous test and
recovers
 the Landau-Yang theorem in the light front reference frame.
 

\section{Gluon radiation process:\\ \boldmath$G_a + N \to (\bar cc)^+_1 + G_b 
+ X$}\label{diagrams}
\setcounter{equation}{0}

Nuclear shadowing of gluons is treated by the parton model as glue-glue
fusion in the infinite momentum frame of the nucleus. On the other hand, in
the rest frame it is expressed in terms of the Glauber like shadowing for the
process of gluon radiation,
 \beq
G_a + N \to (\bar cc)^+_1 + G_b + X\ ,
\label{d0}
 \eeq
 and production of $\bar cc$ pair with positive C-parity in a
color-singlet state. $G_{a,b}$ are gluons in color states $a$ and
$b$. Switching to impact parameter representation one can easily
sum up all multiple scattering corrections for this reaction on a
nucleus, since they take a simple eikonal form \cite{zkl}. Besides,
one can employ the well developed color dipole phenomenology with
parameters fixed by data from DIS.

\begin{figure}[tbh]
\includegraphics{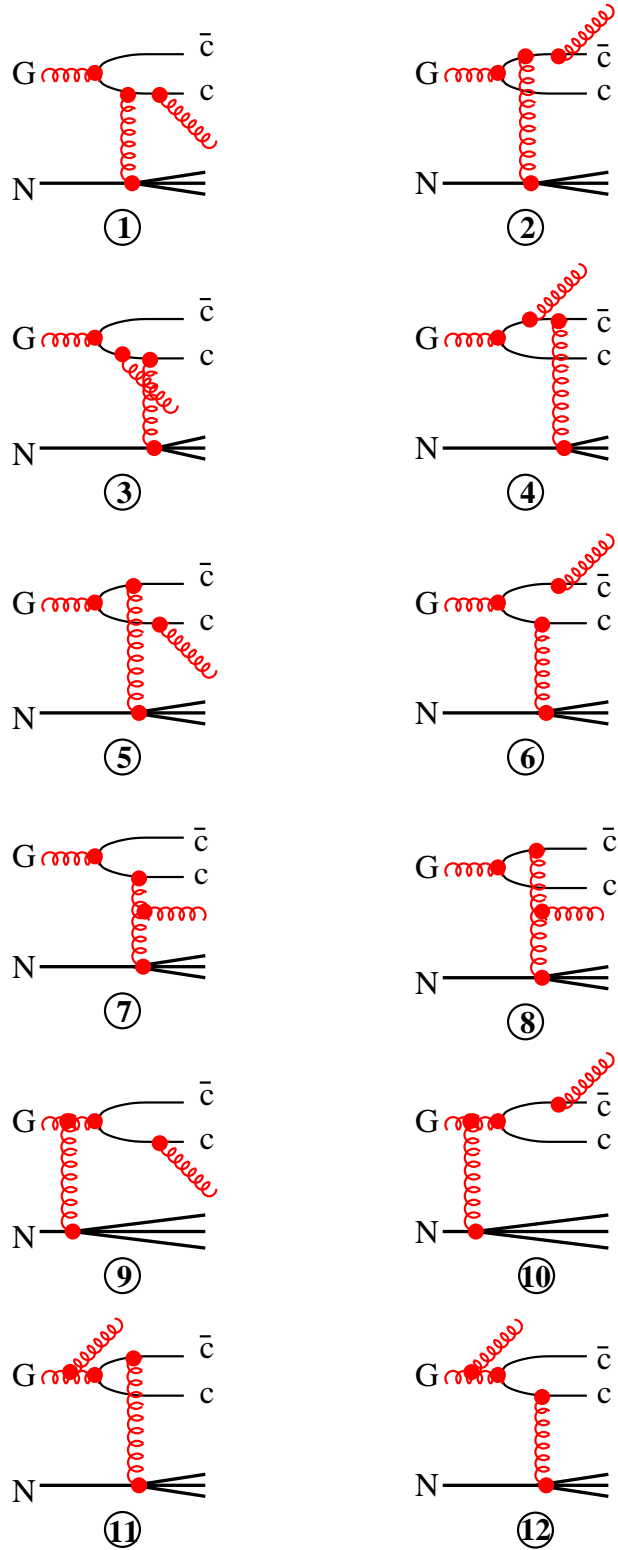}
\begin{center}
\vspace{19.4cm}
\parbox{13cm}
{\caption[Delta]
{\sl Born graphs contributing to $\bar cc$ pair production accompanied with 
radiation of a gluon. The $\bar cc$ is in a color-singlet and C-even state.}
\label{graphs}}
\end{center}
 \end{figure}

The amplitude of the process Eq.~(\ref{d0}) is described in Born
approximation by the set of 12 Feynman graphs depicted in
Fig.~\ref{graphs}. The produced $\bar cc$ pair is assumed to be
colorless and to have positive C-parity.
Note that the two sets of diagrams, 1-10 and 11-12, are gauge invariant
over the t-channel gluon, i.e. the two contribution to the cross section 
are infra-red stable. However, they correspond to different Fock 
components of the incident gluon in  the light-cone approach.
The first set (1-10) corresponds to a fluctuation containing a 
colorless $\bar cc$ and a gluon (see below), white the second one (11-12)
describes a fluctuation with a colored $\bar cc$ and a gluon. In the 
latter case the $\bar cc$ has to interact with the target to become 
colorless, in the same way as for the $G\to\bar cc$ Fock component 
considered above in Sect.~\ref{quarks}. The gluon does not interact at 
all at  the Born level, therefore these graphs 11-12 do not contribute to 
gluon shadowing.

The amplitude of a colorless $\bar cc$ production corresponding to 
graphs 1-10 has the following structure,
 \beq
A^{\bar \mu\mu} = 
\sum\limits_{l=1}^{10} A^{\bar \mu\mu}_l\ ,
\label{d1}
 \eeq
 where
 \beq
A^{\bar \mu\mu}_l = \frac{i\,\alpha_s^{3/2}}
{4\sqrt{3}\,D_l\,(k_T^2+\lambda^2)}\,
\sum\limits_{d=1}^8 f_{abd}\,
F^{(d)}_{GN\to X}(k_T,\{X\})\ 
\xi^\mu_c\Bigr.^\dagger\,\hat\Gamma_l\,
\bar\xi^{\bar\mu}_{\bar c}\ .
\label{d2}
 \eeq
 Here $\lambda$ is the effective gluon mass aimed to incorporate
confinement, its value we discuss later.
The amplitude $F^{(d)}_{GN\to X}(k_T,\{X\})$ determines the 
unintegrated gluon density as it is introduced in (\ref{1.0}),
 \beq
\int d\{X\}\sum\limits_{d=1}^8 
\Bigl|F^{(d)}_{GN\to X}(k_T,\{X\})\Bigr|^2 =
4\pi\,{\cal F}(k_T^2,x)\ .
\label{d2a}
 \eeq
 Here
 \beqn
x &=& \frac{M^2(\bar c,c,G)}{s}\ ;
\label{d2b}\\
M^2(\bar c,c,G) &=&
\frac{m_c^2+k_1^2}{\alpha_1} +
\frac{m_c^2+k_2^2}{\alpha_2} +
\frac{\lambda^2+k_3^2}{\alpha_3}\ ,
\label{d2c}
 \eeqn
where $\vec k_1,\ \vec k_2,\ \vec k_3$ and $\alpha_1,\
\alpha_2,\ \alpha_3$ are the transverse momenta and fractions of
the initial light-cone momentum of the projectile gluon carried by
the produced $\bar c,\ c$ and $G$ (see Fig.~\ref{graphs}),
respectively, and
 \beq
\vec k_T =
\vec k_1 +\vec k_2 +\vec k_3\ .
\label{d2d} 
 \eeq
The $c$-quark spinors $\xi$ are defined in
(\ref{1.02a}); $\{X\}$ is the set of variables describing  the state $X$; 
the 10 vertex functions $\hat\Gamma_l$ read,
 \beqn
\hat\Gamma_1 &=&
\hat U_1(\vec k_1,\alpha_1)\, 
\hat V_1(\vec k_{23},\alpha_2,\alpha_3)
\ ;\nonumber\\
\hat\Gamma_2 &=& - 
\hat V_1(\vec k_{13},\alpha_1,\alpha_3)\,
\hat U_1(\vec k_2,\alpha_2)\, 
\ ;\nonumber\\
\hat\Gamma_3 &=&
\alpha_1\,\hat U_1(\vec k_1,\alpha_1)\, 
\hat V_1(\vec k_{23}-\alpha_3\vec k_T,\alpha_2,\alpha_3)
\ ;\nonumber\\
\hat\Gamma_4 &=& - \alpha_2\,
\hat V_2(\vec k_{13}-\alpha_3\vec k_T,\alpha_1,\alpha_3)\,
\hat U_2(\vec k_2,\alpha_2)\, 
\ ;\nonumber\\
\hat\Gamma_5 &=& \alpha_2\alpha_3\,
\hat U_1(\vec k_1 - \vec k_T,\alpha_1)\, 
\hat V_1(\vec k_{23},\alpha_2,\alpha_3)
\ ;\nonumber\\
\hat\Gamma_6 &=& -\alpha_1\alpha_3\, 
\hat V_2(\vec k_{13},\alpha_1,\alpha_3)\,
\hat U_2(\vec k_2 - \vec k_T,\alpha_2)\, 
\ ;\nonumber\\
\hat\Gamma_7 &=& -2\alpha_1\,
\hat U_1(\vec k_1,\alpha_1)\, 
\hat V_1(\vec k_{23} + \alpha_2\vec k_T,\alpha_2,\alpha_3)
\ ;\nonumber\\
\hat\Gamma_8 &=&  2\alpha_2\,
\hat V_2(\vec k_{13} + \alpha_1\vec k_T,\alpha_1,\alpha_3)\,
\hat U_2(\vec k_2,\alpha_2)\, 
\ ;\nonumber\\
\hat\Gamma_9 &=& -2\alpha_2\alpha_3\,
\hat U_1(\vec k_1 - \alpha_1\vec k_T,\alpha_1)\, 
\hat V_1(\vec k_{23},\alpha_2,\alpha_3)
\ ;\nonumber\\
\hat\Gamma_{10} &=& 2\alpha_1\alpha_3\, 
\hat V_2(\vec k_{13},\alpha_1,\alpha_3)\,
\hat U_2(\vec k_2 - \alpha_2\vec k_T,\alpha_2)\ .
\label{d3}
 \eeqn 
 Here 
 \beqn 
&&\vec k_{13} = 
\alpha_3\vec k_1 - \alpha_1\vec k_3\ ; 
\nonumber\\ 
&&\vec k_{23} =
\alpha_3\vec k_2 - \alpha_2\vec k_3\ ; 
\nonumber\\ 
&&\alpha_1+\alpha_2+\alpha_3 = 1\ . 
\label{d3a}
 \eeqn                                                           

The matrixes $\hat U_{1,2}$ and $\hat V_{1,2}$ are defined as follows,
 \beqn
\hat U_1(\vec k_1,\alpha_1) &=& 
m_c\, \vec\sigma\cdot\vec e +
(1-2\alpha_1)\,\vec e\cdot\vec k_1 + 
i\,(\vec e\times\vec n)\cdot\vec k_1\ ;
\nonumber\\
\hat U_2(\vec k_2,\alpha_2) &=& 
m_c\, \vec\sigma\cdot\vec e +
(1-2\alpha_2)\,\vec e\cdot\vec k_2 + 
i\,(\vec e\times\vec n)\cdot\vec k_2\ ;
\\label{d4a}\\
\hat V_1(\vec k_{23},\alpha_2,\alpha_3) &=&
(2\alpha_2+\alpha_3)\,\vec k_{23}\cdot\vec e +
i\,m_c\,\alpha_3^2\,(\vec n\times\vec e)\cdot\vec\sigma -
i\,\alpha_3\,(\vec k_{23}\times\vec e)\cdot\vec\sigma\ ;
\nonumber\\
\hat V_2(\vec k_{13},\alpha_1,\alpha_3) &=&
(2\alpha_1+\alpha_3)\,\vec k_{13}\cdot\vec e +
i\,m_c\,\alpha_3^2\,(\vec n\times\vec e)\cdot\vec\sigma -
i\,\alpha_3\,(\vec k_{13}\times\vec e)\cdot\vec\sigma\ .
\nonumber\\
\label{d4b}
\eeqn

The functions $D_l$ in the denominator of (\ref{d2}) read,
 \beqn
D_1 &=& \Delta_0(\vec k_1,\alpha_1)\,
\Delta_2(\vec k_{23},\alpha_2,\alpha_3)\ ;
\nonumber\\
D_2 &=& \Delta_0(\vec k_2,\alpha_2)\,
\Delta_2(\vec k_{13},\alpha_1,\alpha_3)\ ;
\nonumber\\
D_3 &=& \Delta_0(\vec k_1,\alpha_1)\,
\Delta_1(\vec k_1,\vec k_{23} - 
\alpha_3\vec k_T,\alpha_1,\alpha_2,\alpha_3)\ ;
\nonumber\\
D_4 &=& \Delta_0(\vec k_2,\alpha_2)\,
\Delta_1(\vec k_2,\vec k_{13} - 
\alpha_3\vec k_T,\alpha_2,\alpha_1,\alpha_3)\ ;
\nonumber\\
D_5 &=& \Delta_1(\vec k_1 - \vec k_T,
\vec k_{23},\alpha_1,\alpha_2,\alpha_3)\,
\Delta_2(\vec k_{23},\alpha_2,\alpha_3)\ ;
\nonumber\\
D_6 &=& \Delta_1(\vec k_2 - \vec k_T,
\vec k_{13},\alpha_2,\alpha_1,\alpha_3)\,
\Delta_2(\vec k_{13},\alpha_1,\alpha_3)\ ;
\nonumber\\
D_7 &=& \Delta_0(\vec k_1,\alpha_1)\,
\Delta_1(\vec k_1,\vec k_{23} + \alpha_2\vec k_T,
\alpha_1,\alpha_2,\alpha_3)\ ;
\nonumber\\
D_8 &=& \Delta_0(\vec k_2,\alpha_2)\,
\Delta_1(\vec k_2,\vec k_{13} + \alpha_1\vec k_T,
\alpha_2,\alpha_1,\alpha_3)\ ;
\nonumber\\
D_9 &=& \Delta_2(\vec k_{23},\alpha_2,\alpha_3)\,
\Delta_1(\vec k_1 - \alpha_1\vec k_T,\vec k_{23},
\alpha_1,\alpha_2,\alpha_3)\ ;
\nonumber\\
D_{10} &=& \Delta_2(\vec k_{13},\alpha_1,\alpha_3)\,
\Delta_1(\vec k_2 - \alpha_2\vec k_T,\vec k_{13},
\alpha_2,\alpha_1,\alpha_3)\ ,
\label{d5}
 \eeqn
 where
 \beqn
\Delta_0(\vec k_1,\alpha_1) &=&
m_c^2 + k_1^2 - \alpha_1(1-\alpha_1)\,\lambda^2\ ;
\nonumber\\
\Delta_2(\vec k_{13},\alpha_1,\alpha_3) &=&
\alpha_3^3\,m_c^2 + 
\alpha_1(\alpha_1+\alpha_3)\,\lambda^2 + 
k_{13}^2\ ;\nonumber\\
\Delta_1(\vec k_1,\vec k_{23},\alpha_1,\alpha_2,\alpha_3) &=& 
(\alpha_1-\alpha_3)(\alpha_3\,m_c^2 + 
\alpha_1\alpha_2\,\lambda^2) +
\alpha_1k_{23}^2 + \alpha_2\alpha_2\,k_1^2\ .
\label{d6}
 \eeqn

Functions $\Delta_0$, $\Delta_1$ and $\Delta_2$ are not independent, but 
satisfy the relation
 \beqn
&&\Bigl[\Delta_0(\vec k_1,\alpha_1)\,
\Delta_2(\vec k_{23},\alpha_2,\alpha_3)\Bigr]^{-1} =
\alpha_1\,\Bigl[\Delta_0(\vec k_1,\alpha_1)\,
\Delta_1(\vec k_1,\vec k_{23},\alpha_1,\alpha_2,\alpha_3)
\Bigr]^{-1}
\nonumber\\ &+&
\alpha_2\alpha_3\,
\Bigl[\Delta_2(\vec k_{23},\alpha_2,\alpha_3)\,
\Delta_1(\vec k_1,\vec k_{23},\alpha_1,\alpha_2,\alpha_3)
\Bigr]^{-1}\ .
\label{d7}
 \eeqn  

The sum of ten amplitudes Eq.~(\ref{d2}) is convenient to split up 
to four terms,
 \beq
\hat M = \sum\limits_{l=1}^{10} = 
\sum\limits_{i=1}^{4}\hat M_i\,
\frac{\hat\Gamma_l}{D_l}\ ,
\label{d8} 
 \eeq
which can be represented using relation (\ref{d7}) 
in the following form,
 \beqn
\hat M_1 &=& 
\alpha_1\,\hat\nu_1(\vec k_1,\alpha_1)\,
\Bigr[\hat\mu_1(\vec k_1,\vec k_{23},\alpha_1,\alpha_2,\alpha_3)
+ \hat\mu_1(\vec k_1,\vec k_{23} - \alpha_3\vec k_T,
\alpha_1,\alpha_2,\alpha_3)\nonumber\\  &-& 
2\,\hat\mu_1(\vec k_1,\vec k_{23} + \alpha_2\vec k_T,
\alpha_1,\alpha_2,\alpha_3)\Bigr] 
\label{d9a}\\
\hat M_2 &=& 
- \alpha_2\,\Bigl[
\hat\mu_2(\vec k_2,\vec
k_{13},\alpha_2,\alpha_1,\alpha_3) + 
\hat\mu_2(\vec k_2,\vec
k_{13} - \alpha_3\vec k_T,\alpha_1,\alpha_2,\alpha_3)
\nonumber\\  &-& 
2\hat\mu_2(\vec k_2,\vec 
k_{13} + \alpha_1\vec k_T,\alpha_1,\alpha_2,\alpha_3)
\Bigr]\,\nu_2(\vec k_2,\alpha_2)\ ;
\label{d9b}\\
\hat M_3 &=& 
\alpha_2\alpha_3\,\Bigr[\hat\lambda_1(\vec k_1,\vec 
k_{23},\alpha_1,\alpha_2,\alpha_3)
+ \hat\lambda_1(\vec k_1-\vec k_T,\vec k_{23},
\alpha_1,\alpha_2,\alpha_3) 
\nonumber\\  &-&
2\hat\lambda_1(\vec k_1-\alpha_1\vec k_T,\vec k_{23}, 
\alpha_1,\alpha_2,\alpha_3)\Bigr]\,
\hat\rho_1(\vec k_{23},\alpha_2,\alpha_3)\ ;
\label{d9c}\\
\hat M_4 &=&
-\alpha_1\alpha_2\,\hat\rho_2(\vec k_{13},\alpha_1,\alpha_3)\,
\Bigl[\hat\lambda_2(\vec k_2,\vec
k_{13},\alpha_1,\alpha_2,\alpha_3) 
\nonumber\\  &+&
\hat\lambda_2(\vec k_2 - \vec k_T,\vec k_{13},
\alpha_1,\alpha_2,\alpha_3)
- 2\hat\lambda_2(\vec k_2 - \alpha_2\vec k_T,\vec 
k_{13},\alpha_1,\alpha_2,\alpha_3)\Bigr]\ .
\label{d9d}
 \eeqn 
 The following notations are used here,
 \beqn
\hat\nu_1(\vec k_1,\alpha_1) &=&
\frac{\hat U_1(\vec k_1,\alpha_1)}
{\Delta_0(\vec k_1,\alpha_1)}\ ;
\nonumber\\
\hat\nu_2(\vec k_2,\alpha_2) &=&
\frac{\hat U_2(\vec k_2,\alpha_2)}
{\Delta_0(\vec k_2,\alpha_2)}\ ;
\label{d10}
 \eeqn
 \beqn
\hat\mu_1(\vec k_1,\vec k_{23},\alpha_1,\alpha_2,\alpha_3)
&=& \frac{\hat V_1(\vec k_{23},\alpha_2,\alpha_3)}
{\Delta_1(\vec k_1,\vec k_{23},\alpha_1,\alpha_2,\alpha_3)}
\ ;\nonumber\\ 
\hat\mu_2(\vec k_2,\vec k_{13},\alpha_1,\alpha_2,\alpha_3)
&=& \frac{\hat V_2(\vec k_{13},\alpha_1,\alpha_3)}
{\Delta_1(\vec k_2,\vec k_{13},\alpha_1,\alpha_2,\alpha_3)}
\ ;
\label{d11}
 \eeqn
 \beqn
\hat\lambda_1(\vec k_1,\vec k_{23},\alpha_1,\alpha_2,\alpha_3)
&=& \frac{\hat U_1(\vec k_1,\alpha_1)}
{\Delta_1(\vec k_1,\vec k_{23},\alpha_1,\alpha_2,\alpha_3)}
\ ;\nonumber\\ 
\hat\lambda_2(\vec k_2,\vec k_{13},\alpha_1,\alpha_2,\alpha_3)
&=& \frac{\hat U_2(\vec k_2,\alpha_2)}
{\Delta_1(\vec k_2,\vec k_{13},\alpha_1,\alpha_2,\alpha_3)}
\ ;
\label{d12}
 \eeqn
 \beqn
\hat\rho_1(\vec k_{23},\alpha_2,\alpha_3) &=&
\frac{\hat V_1(\vec k_{23},\alpha_2,\alpha_3)}
{\Delta_2(\vec k_{23},\alpha_2,\alpha_3)}\ ;
\nonumber\\
\hat\rho_2(\vec k_{13},\alpha_1,\alpha_3) &=&
\frac{\hat V_2(\vec k_{13},\alpha_1,\alpha_3)}
{\Delta_2(\vec k_{13},\alpha_1,\alpha_3)}\ ;
\label{d13}
 \eeqn

Apparently, the amplitude operators in Eqs.~(\ref{d9a})-(\ref{d9d})  
vanish, $\hat M_i \to 0$ when $\vec k_T \to 0$ what guarantees
infra-red stability of the cross section of gluon radiation.

Since the cross section steeply falls down at large
$k_T^2$ and $\hat M_i(k_T=0)=0$ we can expand the amplitudes at 
small $k_T$,
 \beq
M_i(k_T) \approx \vec k_T\left[
\frac{\partial}{\partial \vec k_T}\,
\hat M_i(\vec k_T)\right]_{k_T=0}\ .
\label{d14}
\eeq
 Using Eqs.~(\ref{d9a})-(\ref{d9d}) we find,
 \beqn
\left.\frac{\partial\hat M_1}
{\partial \vec k_T}\right|_{k_T=0} &\approx&
-\alpha_1\,\hat\nu_1\,
\frac{\partial\hat\mu_1}{\partial \vec k_{23}}\,
(\alpha_3+2\,\alpha_2)\ ;
\label{d15a}\\
\left.\frac{\partial\hat M_2}
{\partial \vec k_T}\right|_{k_T=0} &\approx&
\alpha_1\,\frac{\partial\hat\mu_2}
{\partial \vec k_{13}}\,\hat\nu_2\,
(\alpha_3+2\,\alpha_1)\ ;\\
\left.\frac{\partial\hat M_3} 
{\partial \vec k_T}\right|_{k_T=0} &\approx& 
-\alpha_2\alpha_3\,\frac{\partial\hat\lambda_1} 
{\partial \vec k_{2}}\,\hat\rho_1\, 
(1-2\,\alpha_1)\ ;\\
\left.\frac{\partial\hat M_4} 
{\partial \vec k_T}\right|_{k_T=0} &\approx& 
\alpha_1\alpha_3\,\hat\rho_2\,\frac{\partial\hat\lambda_2} 
{\partial \vec k_{2}}\, 
(1-2\,\alpha_2)\ .
\label{d15b}
 \eeqn
 
Using the obvious relations
 \beqn
1-2\alpha_1 &=& 
\alpha_2-\alpha_1+\alpha_3\ ,
\nonumber\\
1-2\alpha_2 &=& \alpha_1-\alpha_2+\alpha_3\ ,
\label{d16}
 \eeqn
 we conclude that the terms $\hat M_3$ and $\hat M_4$ in the
amplitude of $\bar cc$ pair production are negligible compared to
$\hat M_1$ and $\hat M_2$. Indeed, since the $\bar cc$ must be
projected on the wave function of the heavy charmonium, one can
make use of the fact that it is a nonrelativistic system, and 
that the radiated gluon is predominantly soft, i.e.
 \beqn
|\vec k_1 - \vec k_2| &\ll& m_c\ ,
\nonumber\\
|\alpha_1 - \alpha_2| &\ll& 1\ ,
\nonumber\\                                              
\alpha_3 &\ll& 1\ .
\label{d17}
 \eeqn 

These conditions also help to simplify essentially the
Eqs.~(\ref{d15a}) and (\ref{d15b}) for $\hat M_1$ and $\hat M_2$.
First of all, for vanishing $\alpha_3$ one can neglect the terms
containing the Pauli matrixes in the definition (\ref{d4b}) of
$\hat V_1$ and $\hat V_2$ which now commutate with $\hat U_1$ and
$\hat U_2$. 

Further, for small $\alpha_3$ Eq.~(\ref{d3a}) leads to $\vec k_{13}
\approx -\alpha_1\,\vec k_3$, $\vec k_{23}\approx- \alpha_2\,\vec
k_3$. Since from (\ref{d17}) it follows that
$\alpha_1\approx\alpha_2\approx 1/2$ eventually we arrive at $\vec
k_{13} \approx \vec k_{23} \approx -\vec k_3/2$.
Then the propagator $\Delta_1$ can be represented as
 \beqn
\Delta_1 &=& {1\over8}\,(k_3^2+\tau^2) + 
O[\alpha_3(\vec k_1 - \vec k_2)]\ ;\nonumber\\
\tau^2 &=& \lambda^2 + \alpha_3\,M^2\ ;\nonumber\\
M &=& 2m_c \approx M_{\bar cc}\ .
\label{d18}
 \eeqn

The approximations done we arrive to a simplified form of  the 
amplitude,
 \beqn
\hat M &=& 2\,\left[\frac{m_c\,\vec\sigma\cdot\vec e +
i\,(\vec n\times\vec k_1)\cdot\vec e}
{m_c^2+k_1^2} -
\frac{m_c\,\vec\sigma\cdot\vec e 
i\,(\vec n\times\vec k_2)\cdot\vec e}
{m_c^2+k_2^2}\right]\nonumber\\
&\times& \left[\frac{2(\vec k_3 - \vec k_T)\cdot\vec e}
{(\vec k_3-\vec k_T)^2 + \tau^2} - 
\frac{2\,\vec k_3\cdot\vec e}{k_3^2 + \tau^2}\right]\ .
\label{d19}
 \eeqn
 Here we suppressed in the propagator the term 
$\lambda^2/4 \sim \Lambda_{QCD}^2/4$ which is tiny compared to 
$m_c^2$. 

Using Eq.~(\ref{d19}) the amplitude (\ref{d2}) takes the form
 \beq
A^{\bar\mu\mu}_{ab}(\vec k_{12},\vec k_3,\vec k_T,\alpha_3) =
\int d^2b\,d^2s\,d^2r\,
A^{\bar\mu\mu}_{ab}(\vec b,\vec s,\vec r,\alpha_3)
\exp\Bigl[i(\vec b\cdot\vec k_T +
\vec s\cdot\vec k_3 +\vec r\cdot\vec k_{12}\Bigr]\ ,
\label{d20a}
 \eeq
 where
 \beqn
&& A^{\bar\mu\mu}_{ab}(\vec b,\vec s,\vec r,\alpha_3)
\ =\  \frac{i\,\sqrt{3}}{2}\,
\sum\limits_{d=1}^8 f_{abd}\,
\Phi^{\bar\mu\mu}_G(\vec r,\alpha=1/2)\nonumber\\
&\times&\,
\left\{\Phi_{qG}\left(\vec s + \frac{\vec r}{2}\right)\,
\left[\gamma^{(d)}\Bigl(\vec b + \vec s,\{X\}\Bigr) -
\gamma^{(d)}\Bigl(\vec b - 
\frac{\vec r}{2},\{X\}\Bigr)\right]\right.
\nonumber\\ 
&-& \left.\Phi_{qG}\left(\vec s - \frac{\vec r}{2}\right)\,
\left[\gamma^{(d)}\Bigl(\vec b + \vec s,\{X\}\Bigr) -
\gamma^{(d)}\Bigl(\vec b +
\frac{\vec r}{2},\{X\}\Bigr)\right]
\right\}\ .
\label{d20b}
 \eeqn
 Here $\Phi^{\bar\mu\mu}_G(\vec r,\alpha)$ is the distribution 
amplitude for a $\bar cc$ fluctuation of a gluon given by 
Eq.~(\ref{1.02}), and $\Phi_{qG}(\vec s)$ is the distribution 
amplitude for a quark-gluon fluctuation of a quark which has the 
form,
 \beq
\Phi_{qG}(\vec r) =  \frac{2i}{\pi}\,
\sqrt{\frac{\alpha_s}{3}}\,
\vec e\cdot \vec\nabla_r\,
K_0(\tau r)\ .
\label{d21}
 \eeq 

The profile function $\gamma^{(d)}(\vec b,\{X\})$ in
Eq.~(\ref{d20b}) is related to the amplitude $F^{(d)}(\vec
k_T,\{X\})$ by Fourier transform,
 \beq
\gamma^{(d)}(\vec b,\{X\}) =
\frac{\sqrt{\alpha_s}}{2\pi\sqrt{6}}
\int \frac{d^2k_T}{k_T^2+\lambda^2}\,
e^{-i\vec k_T\cdot\vec b}\,
F^{(d)}_{GN\to X}(\vec k_T,\{X\})\ .
\label{d22}
 \eeq 
 This profile function is related to the unintegrated gluon density
${\cal F}(k_T,x)$ and the dipole cross section $\sigma_{\bar
qq}(r,x)$ by the relation,
 \beqn
&& \int d^2b\,d\Gamma_X\,
\sum\limits_{d=1}^8 \left|\gamma^{(d)}(\vec b + \vec r,\{X\}) -
\gamma^{(d)}(\vec b,\{X\})\right|^2 
\nonumber\\
&=& \frac{4\pi}{3}\,\alpha_s\,
\int \frac{d^2k_T}{k_T^2+\lambda^2}\,
\left(1 - e^{i\vec k_T\cdot\vec r}\right)\,
{\cal F}(k_T,x) = \sigma_{\bar qq}(r,x)\ .
\label{d23}
 \eeqn

Now we can calculate the amplitude of $\chi$ production accompanied 
by gluon radiation which corresponds to the graphs depicted in 
Fig.~\ref{graphs} with the $\bar cc$ pair projected to the wave 
function of $\chi$,
 \beq
A^\chi_{ab}(\vec k_T,\vec k_3) =
\sum\limits_{\bar\mu\mu}\,
\int d\alpha\,d^2b\,d^2s\,d^2r\,
\Phi_{\chi}^{\bar\mu\mu}(\vec r,\alpha)\,
A^{\bar\mu\mu}_{ab}(\vec b,\vec s,\vec r)\,
\exp\left[i(\vec b\cdot\vec k_T +
\vec s\cdot\vec k_3)\right]\ .
\label{d24}
 \eeq

Note that the typical value of $r\sim 1/m_c$ is much smaller than 
the mean separation $b\sim 1/\Lambda_{QCD}$. Therefore, we can use 
the approximation $\gamma^{(d)}(\vec b \pm \vec r,\{X\}) \approx 
\gamma^{(d)}(\vec b,\{X\})$ in Eq.~(\ref{d20b}), then the amplitude 
$A^{\bar\mu\mu}$ takes the form which is known for the radiation of 
a heavy photon by a quark, i.e. the amplitude of Drell-Yan process 
\cite{hir,kst1},
 \beqn
A^\chi_{ab}(\vec k_T,\vec k_3,\alpha_3) &\approx&
-\frac{i\sqrt{3}}{2}\,\sum\limits_{\bar\mu\mu}\,
\int d^2b\,d^2s\, \Psi(s,\alpha_3)\,
\exp\left[i(\vec b\cdot\vec k_T +
\vec s\cdot\vec k_3)\right]
\nonumber\\ &\times&
\left[\gamma^{(d)}\Bigl(\vec b + \vec s,\{X\}\Bigr) -
\gamma^{(d)}\Bigl(\vec b,\{X\}\Bigr)\right]\ ,
\label{d25}
 \eeqn
 where $\Psi(s,\alpha_3)$ defined in (\ref{3.3}) is the effective
distribution amplitude for s $\chi-G$ fluctuation of a gluon, which is
the analog for the $\gamma^*\,q$ fluctuation of a quark.

The two terms in square brackets in (\ref{d25}) correspond to the two
diagrams in Fig.~\ref{dy} which are the same as for Drell-Yan reaction
if to replace the virtual photon by $\chi$ and the incoming and recoil 
quarks by gluons. Therefore, it is natural to expect the corresponding 
cross section to have the same factorized form as for Drell-Yan 
reaction. Indeed, squaring expression Eq.~(\ref{d25}) and using 
Eq.~(\ref{d23}) we arrive at the expression Eq.~(\ref{3.2}) for the 
cross section.

\end{document}